\documentclass{article}

\usepackage{xspace}
\usepackage{paralist}
\usepackage{subfig}
\usepackage[utf8]{inputenc}
\usepackage{amsmath}
\usepackage{amssymb}
\usepackage{float}
\usepackage{graphicx}
\usepackage{wrapfig}
\usepackage[basic]{complexity}
\usepackage{xspace}
\usepackage[table]{xcolor}
\usepackage{hyperref}
\usepackage{array}
\usepackage{placeins}
\usepackage{geometry}
\usepackage{authblk}

\usepackage[left,pagewise] {lineno}

\usepackage[vlined,linesnumbered]{algorithm2e}
\DontPrintSemicolon

\usepackage[disable]{todonotes}

\usepackage{url}
\usepackage{booktabs}

\newcommand{\martin}[1]{\todo[color=yellow!40]{MN: #1}}

\definecolor{insig-bg}{rgb}{1.0, 0.8, 0.8}
\definecolor{sig-bg}{rgb}{0.8, 1.0, 0.8}

\newcommand{\DA}{\ensuremath{\mathrm{D}_\mathrm{U}}}
\newcommand{\DB}{\ensuremath{\mathrm{D}_{3}}}
\newcommand{\DC}{\ensuremath{\mathrm{D}_{10}}}

\newcommand{\LPO}{\ensuremath{po}\xspace}
\newcommand{\LS}{\ensuremath{s}\xspace}
\newcommand{\LDO}{\ensuremath{do}\xspace}
\newcommand{\LOPO}{\ensuremath{opo}\xspace}

\newcommand{\GPO}{\ensuremath{\mathcal{PO}}\xspace}
\newcommand{\GS}{\ensuremath{\mathcal{S}}\xspace}
\newcommand{\GDO}{\ensuremath{\mathcal{DO}}\xspace}
\newcommand{\GOPO}{\ensuremath{\mathcal{OPO}}\xspace}

\newcommand{\TASKA}{\ensuremath{\mathrm{T}_{\mathrm{S}}}\xspace}
\newcommand{\TASKB}{\ensuremath{\mathrm{T}_{\mathrm{L}}}\xspace}

\definecolor {infocolor} {rgb} {0.6,0.6,0.6}

\title{On the Readability of Boundary Labeling}

\author{Lukas Barth}
\author{Andreas Gemsa}
\author{Benjamin Niedermann}
\affil{Institute of Theoretical Informatics, Karlsruhe Institute of Technology, Germany}
\author{Martin N\"ollenburg}
\affil{Algorithms and Complexity Group, TU Wien, Vienna, Austria}
\date{}

\begin{document}
\maketitle

\begin{abstract}
  Boundary labeling deals with annotating features in images such
  that labels are placed outside of the image and are connected by
  curves (so-called leaders) to the corresponding features.
  While boundary labeling has been extensively investigated from an
  algorithmic perspective, the research on its readability has been
  neglected.  In this paper we present the first formal user study on the
  readability of boundary labeling. We consider the four most studied
  leader types with respect to their performance, i.e., whether and
  how fast a viewer can assign a feature to its label and vice versa.
  We give a detailed analysis of the results regarding
  the readability of the four models and discuss
  their aesthetic qualities based on the users' preference judgments and interviews.
\end{abstract}

\section{Introduction}
Creating complex, but comprehensible figures such as maps,
scientific illustrations, and information graphics is a challenging task comprising multiple design and layout steps. One of these steps is labeling the content of the figure
appropriately. A good labeling conveys information about the
figure without distracting the viewer. It is unintrusive and does not destroy the figure's aesthetics. At the same time it enables the viewer to quickly
and correctly obtain additional information that is not inherently
contained in the figure. Typically multiple features are
labeled by a set of (textual) descriptions called \emph{labels}.
Morrison~\cite{m-ctcc-80}
estimates the time needed for labeling a map to be over 50\% of the
total time when creating a map by hand. Hence, a lot of research
efforts have been made to design algorithms that automate the process of label placement.

To obtain a clear relation between a feature and its label, the label
is often placed closely to it. 
However, in some applications this \emph{internal} labeling is not
sufficient, because either features are densely distributed and there are too many labels to be placed or
any extensive occlusion of the figure's details should be
avoided. While in the first case one may exclude less important labels,
in the second case even a small number of labels may destroy the
readability of the figure. In either case graphic designers often choose to place the
labels outside of the figure and connect the features with their
labels by thin curves, so called \emph{leaders}. This kind of labeling
is commonly found in highly detailed scientific figures as they are
used for example in atlases of human anatomy.  In the graph drawing
community this kind of \emph{external} labeling became well known as
\emph{boundary labeling}. Since Bekos et
al.~\cite{bksw-blmea-07} have introduced boundary
labeling to the graph drawing community, a huge variety of models for
boundary labeling have been considered from an algorithmic
perspective. However, they have not been studied concerning their
readability from a user's perspective. Here we present the first formal user study on the readability of the four most common boundary labeling models.

\begin{figure}[tb]
  \centering
  \includegraphics[scale=0.95]{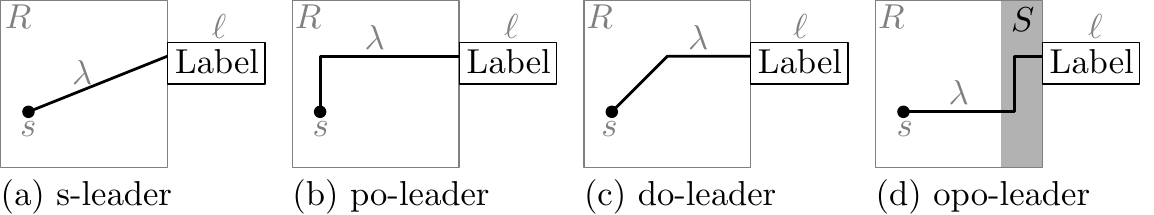}
  \caption{Illustration of leader types. Type-$\LOPO$ leaders use a track routing area $S$.}
  \label{fig:leader-types}
\end{figure}

\paragraph{Models of Boundary Labeling.} The problem of boundary
labeling is formalized as follows (refer to Fig.~\ref{fig:leader-types}).  We are given a rectangle $R$ of height~$h$ and
width~$w$ and a finite set~$P$ of points in~$R$, which we
call~\emph{sites}. Each site~$s$ is assigned to a text that
describes the site. Following traditional map labeling, not the text
itself is considered, but its shape is approximated by its axis-aligned
bounding box~$\ell$. We call~$\ell$ the \emph{label} of the site~$s$. The set of all labels is denoted by~$L$.

The boundary labeling problem then asks for the placement of
labels such that
\begin{inparaenum}[(1)]
\item each label~$\ell\in L$ lies outside of~$R$ and touches the
  boundary of~$R$,
\item no two labels overlap, and
\item for each site~$s$ and its label~$\ell$ there is a
  self-intersection-free curve~$\lambda$ in $R$ that starts at~$s$ and
  ends on the boundary of~$\ell$.
\end{inparaenum} We call the curve~$\lambda$ the \emph{leader} of the
site~$s$ and its label~$\ell$. The end point of~$\lambda$ that
touches~$\ell$ is called the \emph{port} of~$\ell$. Typically, four
main parameters, in which the models differ, are distinguished. The
\emph{label position} specifies on which sides of~$R$ the labels are
placed. The \emph{label size} may be uniform or individually defined
for each label. The \emph{port type} specifies whether \emph{fixed
  ports} or \emph{sliding ports} are used, i.e., whether the position
of a port on its label is pre-defined or flexible. Finally, the
\emph{leader type} restricts the shape of the leaders. As the leader
type is the most distinctive feature of the different boundary
labeling models in the literature, we examine how this parameter
influences the readability. Regarding the other parameters we restrict our attention to one-sided
instances whose labels have unit height, lie on the right side of~$R$
and have fixed ports. In the
following we list the leader types that are most commonly found in the
literature.

Let~$\lambda$ be a leader connecting a site~$s\in P$ with
a label~$\ell \in L$, and let $r$ be the side of~$R$ that is touched
by~$\ell$.
An $\LS$-leader consists of a single straight~($\LS$) line
segment; see Fig.~\ref{fig:leader-types}(a). A~$\LPO$-leader consists of two line segments, the first, starting at~$s$, is parallel ($\mathrm{p}$) to~$r$ and
the second segment is orthogonal ($\mathrm{o}$) to~$r$; see Fig.~\ref{fig:leader-types}(b). A $\LDO$-leader
consists of two line segments, the first, starting at~$s$, is diagonal~($\mathrm{d}$) at some angle $\alpha$ (typically $\alpha=45^\circ$) relative to $r$ and the second segment is
orthogonal ($\mathrm{o}$) to~$r$; see Fig.~\ref{fig:leader-types}(c). An $\LOPO$-leader consists of three line
segments, the first, starting at~$s$, is orthogonal~($\mathrm{o}$) to~$r$, the second is parallel ($\mathrm{p}$) to~$r$, and the
third segment is orthogonal~($\mathrm{o}$) to~$r$; see Fig.~\ref{fig:leader-types}(d).
In case that $\LOPO$-leaders are considered, each leader has its two bends
in a strip~$S$ next to~$r$ whose width is large enough to accommodate all leaders with a minimum pairwise distance of the $p$-segments.
The strip $S$ is
called the \emph{track-routing area} of~$R$.
In the remainder of this
paper, we call a labeling based on $\LS/\LPO/\LDO/\LOPO$-leaders an
$\LS/\LPO/\LDO/\LOPO$-labeling. 

Following Tufte's minimum-ink principle~\cite{t-vdqi-01}, the most common objective in boundary labeling
is to minimize the total leader length, which means minimizing the total overlay of leaders with the given figure. Further, to increase
readability of the labelings, all models usually require that no two
leaders cross each other.
%


\paragraph{Related Work.} \label{sec:related-work} The algorithmic
problem of boundary labeling was introduced at GD 2004 by Bekos et
al.~\cite{bksw-blmea-07}.  They presented efficient algorithms for
models based on~$\LPO$-, $\LOPO$- and $\LS$-leaders. As objective
functions they considered minimizing the number of bends and the total
leader length. While for $\LOPO$-leaders the labels may lie on one,
two, or four sides of~$R$, the labels for~$\LPO$-leaders may lie only
on one or on two opposite sides of~$R$.  In 2005 based on a manual analysis of
hand-drawn illustrations (e.g., anatomic atlases),  Ali et
al.~\cite{ahs-lli3di-05} introduced criteria for boundary labeling
concerning readability, ambiguity and aesthetics. Based on these they
presented force-based heuristics for labeling figures using
$\LS$-leaders and $\LPO$-leaders.  In 2006 Bekos et al.\ considered
$\LOPO$-labelings such that labels appear in multiple stacks
besides~$R$~\cite{bkps-msblp-06}.  Boundary labeling
using~$\LDO$-leaders has been introduced by Benkert et
al.~\cite{bhkn-amcbl-09} in 2009. They investigated algorithms
minimizing a general badness function on $\LDO$- and $\LPO$-leaders
and, furthermore, gave more efficient algorithms for the case that the
total leader length is minimized. In 2010 Bekos et
al.~\cite{bkns-blol-10} presented further algorithms for
$\LDO$-leaders and similarly shaped leaders. Further, Bekos et
al.~\cite{bkps-afbl-10} considered $\LOPO$-labelings such that the
sites may \emph{float} within predefined polygons in~$R$. Nöllenburg
et al.~\cite{nps-d1sbl-10} considered $\LPO$-labelings for a setting
that supports interactive zooming and panning.  In 2011 Gemsa et
al.~\cite{ghn-blapi-11} studied the labeling of panorama images using
vertical~$\LS$-leaders.  Leaders based on Bezi{\'e}r curves and
$\LS$-leaders are further considered in the context of labeling focus
regions by Fink et al.~\cite{fhssw-alfr-12} (2012).  Further, in 2013
Kindermann et al.~\cite{knrssw-2sblas} considered $\LPO$-labelings for
the cases that the labels lie on two adjacent sides, or on more than
two sides.  In 2014 Huang et al.~\cite{hpl-blflp-14} investigated
$\LOPO$-labelings with flexible label positions.

Boundary labeling has also been combined in a \emph{mixed model} with
internal labels, i.e., labels that are placed next to the sites
\cite{ln-sbolb-10,bkps-ctmlbl-11,lns-mml-15}.  \emph{Many-to-one}
boundary labeling is a further variant, where each label may connect
to multiple
sites~\cite{lky-m1bl-08,l-cfm1blh-10,bcf-m1blb-13}. Finally, boundary
labeling has also been considered in the context of \emph{text
  annotations} \cite{lwy-blta-09,klw-lblat-14}.

\newcommand{\XSPACE}{\quad}
\begin{table}[]
\centering
\caption{Summary of related work broken down into the considered leader types. The considered leader types are marked by $\times$. If a natural extension of a leader type has been investigated, then it is marked by $\star$. }
\label{table:related-work}
\begin{tabular}{llccccc}
\toprule
\multicolumn{1}{c}{\textbf{Year}\XSPACE\XSPACE} & \multicolumn{1}{l}{\textbf{Reference}\XSPACE} & \multicolumn{5}{c}{\textbf{Leader Type}}                                                                                   \\\cmidrule(lr){3-7}
\multicolumn{1}{c}{}      & \multicolumn{1}{c}{}     & \multicolumn{1}{c}{\XSPACE$\LS$\XSPACE} & \multicolumn{1}{c}{\XSPACE$\LPO$\XSPACE} & \multicolumn{1}{c}{\XSPACE$\LDO$\XSPACE} & \multicolumn{1}{c}{\XSPACE$\LOPO$\XSPACE} & \multicolumn{1}{c}{\XSPACE other \XSPACE} \\
\midrule
  2004 \XSPACE    &Bekos et al.\cite{bksw-blmea-07}                    & \XSPACE$\times$\XSPACE                  & \XSPACE$\times$\XSPACE                   &                            & \XSPACE$\times$\XSPACE        &            \\
\midrule
  2005\XSPACE    &Ali et al.\cite{ahs-lli3di-05}                     & \XSPACE$\times$\XSPACE                  & \XSPACE$\times$\XSPACE                   &                            &   &                          \\\midrule
  2006\XSPACE    &Bekos et al.\cite{bkps-msblp-06}                     &                           &                            &                            & \XSPACE$\times$\XSPACE       &             \\\midrule
  2008\XSPACE     & Lin et al.\cite{lky-m1bl-08}                     &                           &      \XSPACE$\star$\XSPACE                      &                            &   \XSPACE$\star$\XSPACE      &                    \\\midrule
  2009 \XSPACE   & Benkert et al.\cite{bhkn-amcbl-09}                     &                           & \XSPACE$\times$\XSPACE                   & \XSPACE$\times$\XSPACE   &                &                             \\
 \XSPACE  \XSPACE     & Lin et al.\cite{lwy-blta-09}                     &                           &                            &                            &   \XSPACE$\times$\XSPACE       &   \XSPACE$\times$\XSPACE                \\\midrule
  2010 \XSPACE    & Bekos et al.\cite{bkns-blol-10}                    &                           &                            & \XSPACE$\times$\XSPACE                   &                          &   \XSPACE$\times$\XSPACE  \\
 \XSPACE  \XSPACE    & Bekos et al.\cite{bkps-afbl-10}                    &                           &                            &                            & \XSPACE$\times$\XSPACE                  &  \\
  \XSPACE \XSPACE    &Lin \cite{l-cfm1blh-10}                     &                           &                            &                            &       \XSPACE$\star$\XSPACE   &                   \\
\XSPACE \XSPACE      & Löffler and Nöllenburg\cite{ln-sbolb-10}                     &                           &            \XSPACE$\times$\XSPACE                &                            &          &                   \\
\XSPACE  \XSPACE     & Nöllenburg et al.\cite{nps-d1sbl-10}                    &                           & \XSPACE$\times$\XSPACE                   &                            &                             & \\\midrule
  2011 \XSPACE  & Bekos et al.\cite{bkps-ctmlbl-11}                     &                           &                            &                            &    \XSPACE$\times$\XSPACE       &                  \\
 \XSPACE \XSPACE    & Gemsa et al.\cite{ghn-blapi-11}                     & \XSPACE$\times$\XSPACE                  &                            &                            &          &                   \\\midrule
  2012\XSPACE   & Fink et al.\cite{fhssw-alfr-12}                     & \XSPACE$\times$\XSPACE                  &                            &                            &            &   \XSPACE$\times$\XSPACE   \\\midrule
  2013\XSPACE    & Bekos et al.\cite{bcf-m1blb-13}                     &                           &               \XSPACE$\star$\XSPACE             &                            &             &                \\
 \XSPACE \XSPACE   & Kindermann et al.\cite{knrssw-2sblas}                     &                           & \XSPACE$\times$\XSPACE                   &                            &                             & \\\midrule
  2014\XSPACE    & Huang et al.\cite{hpl-blflp-14}                     &                           &                            &                            & \XSPACE$\times$\XSPACE                 &   \\
 \XSPACE \XSPACE    & Kindermann et al.\cite{klw-lblat-14}                     &        \XSPACE$\times$\XSPACE                   &      \XSPACE$\times$\XSPACE                      &                            &       \XSPACE$\times$\XSPACE          &   \XSPACE$\times$\XSPACE         \\\midrule
  2015\XSPACE      & Löffler et al.\cite{lns-mml-15}                     &                           &                            &          \XSPACE$\times$\XSPACE                  &                  &        \\
\midrule
$\sum$&& \XSPACE 5\XSPACE & \XSPACE 9\XSPACE & \XSPACE 3\XSPACE & \XSPACE 9\XSPACE& \XSPACE 4\XSPACE \\
\bottomrule
\end{tabular}
\end{table}

In total we found three papers studying
$\LDO$-leaders, nine studying $\LOPO$-leaders, nine
studying~$\LPO$-leaders, and five papers studying $\LS$-leaders; see Table~\ref{table:related-work}.

\paragraph{Our Contribution.} While boundary labeling has been
extensively investigated from an algorithmic perspective, the research
on the readability of the introduced models has been
neglected. There exist several user studies on the readability and aesthetics of graph drawings. For example Ware et al.~\cite{wpcm-cmga-02}
studied how people perceive links in node-links diagrams. However, to the
best of our knowledge, there are no studies on the readability of any
boundary labeling models. In this paper we present the first user
study on readability aspects of boundary labeling. When reading a
boundary labeling the viewer typically wants to find for a given site
its corresponding label, or vice versa. Hence, a well readable
labeling must facilitate this basic two-way task such that it can be
performed fast and correctly. We call this the \emph{assignment
  task}. In this paper we investigate the assignment task with respect
to the four most established models, namely models using~$\LS$-,
$\LPO$-, $\LOPO$- and $\LDO$-leaders, respectively. To keep the
  number of parameters small, we refrained from considering other
  types of leaders. We conducted a controlled user study with 31
subjects. Further, we interviewed eight participants about their
personal assessment of the leader types.  We obtained the
following main results.
\begin{compactitem}
\item Type-$\LOPO$ leaders lag behind the other leader types in all considered aspects.
\item In the assignment task, $\LDO$-, $\LPO$- and $\LS$-leaders have similar error rates, but $\LPO$-leaders have significantly faster response times than~$\LDO$- and $\LS$-leaders.
\item The participants prefer the leader types in the order $\LDO$, $\LPO$, $\LS$ and $\LOPO$.
\end{compactitem}

\section{Research Questions}\label{sec:hypo}

As argued before, a well readable boundary labeling must allow the viewer
to quickly and correctly assign a label to its site and vice versa. More
specifically, the leader~$\lambda$ connecting the label with its site
must be easily traceable by a human.
We hypothesize that both the response time and the error rate of the
assignment task significantly depend on other leaders running close to and parallel to~$\lambda$ in the following sense.
  \emph{The more parallel segments closely
    surround~$\lambda$, the more the response time and the error rate
    of the assignment task increase.}

However, we did not directly investigate this hypothesis, but we
derived from it two more concrete hypotheses that are based on the
four leader types. These were then investigated in the user study. To
that end, we additionally observe, that in medical figures the density
of the sites varies. Both may occur, figures containing a \emph{dense
  set} of sites, where the sites are placed closely to each other, and
figures containing a \emph{sparse set} of sites, where the sites are
dispersed. We now motivate the hypothesis as follows.

By definition of the models, the number of parallel leader segments in
$\LDO$-, $\LPO$- and $\LOPO$-labelings is linear in the number of
labels per leader, because each pair of leaders has at least one pair of parallel
segments. For~$\LOPO$-labelings each pair of leaders even has up to
three pairs of parallel segments. Additionally, the spacing of the
first orthogonal segments of $\LOPO$-leaders is determined by the
$y$-coordinates of the sites rather than by the (more regularly
spaced) $y$-coordinates of the label ports as in $\LPO$- and
$\LDO$-labelings. In contrast, in an $\LS$-labeling the leaders
typically have different slopes, so that (almost) no parallel line
segments occur. In fact, it is known that the human eye can
distinguish angular differences as small as $10'' \approx
0.003^\circ$~\cite{w-ivpd-12}.
Hence, leaders of $\LDO$-, $\LPO$- and $\LOPO$-labelings, in
particular for a dense set of sites, are closely surrounded
by parallel segments, while~$\LS$-leaders for such a set have very
different slopes. We therefore propose the next hypothesis.
\begin{enumerate}
\item[\emph{(H1)}] \emph{For instances containing a dense set of
    sites,
    \begin{compactenum}[(a)]
    \item the assignment task on $\LS$-labelings has a significantly
      smaller response time and error rate than on $\LDO$-,
      $\LPO$-, and $\LOPO$-labelings.
    \item the assignment task on $\LDO$- and $\LPO$-labelings
      has a significantly smaller response time and error rate than
      on $\LOPO$-labelings.
    \end{compactenum}
}
\end{enumerate}

Considering a sparse set of  sites, $\LDO$- and $\LPO$-labelings
still have many parallel line segments, but this time they are more
dispersed. This is normally not true for $\LOPO$-leaders because the
actual routing of those leaders occurs in a thin routing area at the
boundary of~$R$. Hence, we propose the next hypothesis.
\begin{enumerate}
\item[\emph{(H2)}] \emph{For instances containing a sparse set of sites, the assignment task on $\LOPO$-labelings has a significantly greater response time and error rate than on $\LDO$-, $\LPO$-, and $\LS$-labelings.}
\end{enumerate}

In summary, we expect that $\LOPO$-labelings perform
worse than the other three, that $\LDO$- and $\LPO$-labelings perform similar, and that $\LS$-labelings perform best.


\section{Design of the Experiment}\label{sec:desgin}
This section presents the tasks, the stimuli, and the experimental procedure that we used to conduct the user study.

\paragraph{Tasks.}
In order to test our hypotheses we presented instances of boundary labeling to the participants and asked them to perform the following two tasks.
\begin{compactenum}
\item Label-\textbf{S}ite-Assignment ($\TASKA$): In an instance containing a
  highlighted label select the related site.
\item Site-\textbf{L}abel-Assignment ($\TASKB$): In an instance containing a
  highlighted site select the related label.
\end{compactenum}

\begin{figure}[tb]
  \centering
  \includegraphics[page=3]{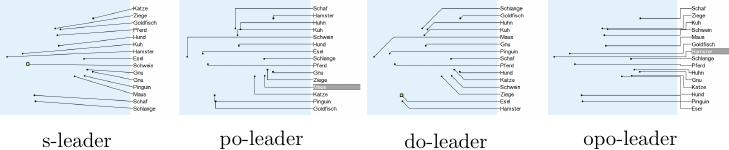}
  \caption{Examples of stimuli for both tasks and all four leader types.}
  \label{fig:example-stimuli}
\end{figure}

\paragraph{Stimuli.}
The stimuli are automatically generated boundary
labelings, each using the same basic drawing style.
In order to remove confounding effects between background image and leaders we use a plain light blue background. Points, leaders and label texts are drawn in the same style and in black color. Highlighted points are drawn as slightly larger yellow-filled squares with black boundary rather than small black disks. Highlighted labels are shown as white text on a dark gray background. Figure~\ref{fig:example-stimuli} exemplarily shows four stimuli.


For all instances we defined~$R$ to be a rectangle of~$500\times 750$
pixels. In addition to the four leader types as the main factor of interest, we identified three secondary factors that may have an impact on the resulting labelings. This yields four parameters to classify an instance.
The first parameter is the \emph{number} $\mathcal N=\{15,30\}$ of
sites that are contained in the instance. We have chosen $15$ sites to
obtain small instances and $30$ sites to obtain large instances, which
are typical numbers, e.g., for medical drawings.
The second parameter is the \emph{distribution}~$\mathcal D=\{\DA,\DB,\DC\}$
that is used for randomly placing the sites in~$R$.  We define $\DA$ to be a
uniform distribution.  This distribution yields instances whose sites
are dispersed in~$R$ without having a certain
spatial structure. 
However,
considering, e.g., medical drawings, the instances often consist of
spatial clusters. We model such a single cluster by a normal
distribution. More precisely, we define~$\DB$
and $\DC$ to be normal distributions with mean~$\mu=(250, 375)$ at the center of $R$, and variance
$\sigma=3000$ and $\sigma=10000$ in both directions, respectively. Hence,~$\DB$ yields
instances consisting of a dense set of sites, while~$\DC$ yields
instances consisting of a sparse set of sites. 
In order to avoid
cluttered sets of sites and degenerated instances, where sites lie too
close  to the boundary of~$R$, we rejected instances where any two
sites have less than~$10$ pixels distance or where a site has less
than~$30$ pixels distance to the boundary of~$R$.
The third parameter is the applied \emph{leader
  type}~$\mathcal T=\{\LDO,\LOPO,\LPO,\LS\}$ as defined above.
Finally, the fourth parameter $\mathcal R=\{0.3,0.6,0.9\}$ can be seen
as a difficulty level and specifies which leader of the instance
should be selected for the tasks~\TASKA~and \TASKB. This is
accomplished by scoring each leader with respect to how much ink is
close to it in the drawing. More specifically, ranking a
leader~$\lambda_i$ is done as follows: For every other
leader~$\lambda_j$, points are linearly sampled on~$\lambda_j$ with
one pixel distance from each other. For each such point, the minimum
distance~$d$ to~$\lambda_i$ is computed.  Then, every sample point
contributes~$\frac{1}{d^2}$ to the \emph{ink score} of~$\lambda_i$. The
parameter~$r\in \mathcal R$ then selects the leader~$\lambda$ whose ink score is
the~$r$-quantile among the ink scores of all leaders in the instance. Thus the parameter~$\mathcal R$ lets us control the relative difficulty of the chosen leader.

The parameter space $\mathcal N\times \mathcal D \times \mathcal T \times
\mathcal R$ gives us the possibility to cover a large variety of
different instances.
For each of the 72 possible choices of parameters $(n,d,t,r) \in \mathcal N\times \mathcal D
\times \mathcal T \times \mathcal R$ we have generated
two instances~$I_1$ and $I_2$, one for each task.  To that end, we used the property that
a leader of any of the four types is uniquely determined by the
location of its port and site. Hence, using integer linear programming (ILP)
we computed a length minimal labeling of the chosen leader type such that
the labels are placed to the right side of~$R$ using one of $150$ equally spaced
ports each. The ILP further ensures that the subset of chosen ports does not create overlapping labels and that no two leaders cross.
In each instance each label is randomly chosen from a pre-defined set of German animal names. For $\LOPO$-labelings, the
track routing area and the routing of the leaders is chosen such that the
$p$-segments of any two leaders have horizontal distance of at least~$10$
pixels from each other. 

It will occur in the instances that leaders lie closely together, e.g., see
$\LOPO$-labeling in Fig.~\ref{fig:example-stimuli}. However, we do not
enforce minimum spacing between leaders because neither any of the
studied models nor any of the discussed algorithms enforce
minimum spacing explicitly. In fact, a fixed minimum leader spacing may even lead to infeasible instances for certain leader types.

\paragraph{Procedure.}
The study was run as a within-subject experiment. Four experimental sessions were held in our computer lab at controlled lighting with 12 identical machines and screens using a digital questionnaire in German language.
After agreeing to a consent form, each participant first completed a tutorial
explaining him or her the tasks $\TASKA$ and $\TASKB$ on four
instances, each containing one of the four labeling
types. Participants were instructed to answer the questions as quickly and as accurately as possible.
Afterwards, the actual study started presenting the 144
stimuli to the participant one at a time.  Each stimulus was
revealed to the participant, after he or she clicked a button in the
center of the screen using the mouse. Hence, at the beginning of each
task the mouse pointer was always located at the same position. Then he or
she performed the task by selecting a label or site using the mouse.

The stimuli were divided into 12 blocks consisting of 12 stimuli
each. Each block either contained stimuli only for $\TASKA$ or only for
$\TASKB$. For each participant the stimuli were in random order, but
in alternating blocks, i.e., after completing a block for $\TASKA$ a
block for $\TASKB$ was presented, and vice versa. Between two successive blocks a pause screen stated the task for the next block and participants were asked to take a break of at least 15 seconds before continuing.

Especially for professional printings, e.g. for atlases of human
  anatomy, not only the figure's readability, but also its aesthetics
  is seen to be of great importance. Further, assigning a label to its
  site (or vice versa), the viewer should be able to assess whether he
  or she has done this correctly.  We therefore asked all
participants about their personal assessment of the aesthetics and readability of the leader types after completing the 144
performance trials.  We presented the same four selected instances of
the four leader types to each participant.  To that end, we selected
an instance for each leader type~$t\in \mathcal T$ based on the 144
instances generated for the tasks~$\TASKA$ and $\TASKB$. We score each
instance by the sum of its leaders' ink scores.  Among all instances
with leader type~$t\in \mathcal T$ and 15 sites, we selected the
median instance~$I$ with respect to the instance scores of that
subset.
Hence, for each type of leader we obtain a
moderate instance with respect to our difficulty measure; see Fig.~\ref{fig:stimuli:questionnair} in the appendix.
Each participant was
asked to rate the different leader types using German school grades on a scale from 1 (excellent) to 6 (insufficient), where grades 5 and 6 are both fail-grades, \martin{wenn platz ist alle noten ausschreiben} by answering the following
questions.
\begin{compactenum}[Q1.]
\item How do you rate the appearance of the leader types?
\item For a highlighted site, how easy is it for you to find the
  corresponding label?
\item For a highlighted label, how easy is it for you to find the
  corresponding site?
\end{compactenum}

We further conducted interviews with eight participants
after the experiment, in which they justified their grading.

\section{Results}\label{sec:result}
In total 31 students of computer science in the age between 20 and 30 years completed the experiment, six of them were female and 25
were male. We also asked whether they have fundamental knowledge about labeling figures and maps, which was affirmed by only two participants.

\subsection{Performance Analysis}
\label{sec:analysis-user-studie}
For each of the 144 trials we recorded both the response time and the
correctness of the answer, which allows for analyzing two separate
quantitative performance measures\footnote{Raw data at
  \url{http://i11www.iti.uni-karlsruhe.de/projects/bl-userstudy}}. Response
times were measured from the time a stimulus was revealed until the
participant clicks to give the answer. Response times are normalized
per participant by his/her median response time to compensate for
different reaction times among participants.  We split the data into
four groups by leader type, and call them \GDO, \GPO, \GS~and \GOPO,
respectively. \martin{falls noch Zeit ist: die Reihenfolge von vorne
  ist S, PO, DO, OPO; wollten wir das nicht durchs PAper so
  durchziehen? Ist aber nur ein nice-to-have, spätestens für die final
  version.}

We applied repeated-measures Friedman tests with post-hoc
Dunn-Bonferroni pairwise comparisons in
SPSS\footnote{\url{http://www-01.ibm.com/software/analytics/spss/}} between
the four groups to find significant differences in the performance
data at a significance level of $p=0.05$. We chose a non-parametric
test since our data are not normally distributed. We report the
detailed test results in Table~\ref{table:sig-timing} (response times)
and Table~\ref{table:sig-correctness} (success rates) in the appendix and summarize the main
findings in the following paragraphs.

\begin{figure}[tb]
  \centering
   \subfloat[Normalized response times (logarithmic scale). Smaller values are better than higher values.]{
       \includegraphics[page=1,scale=0.95]{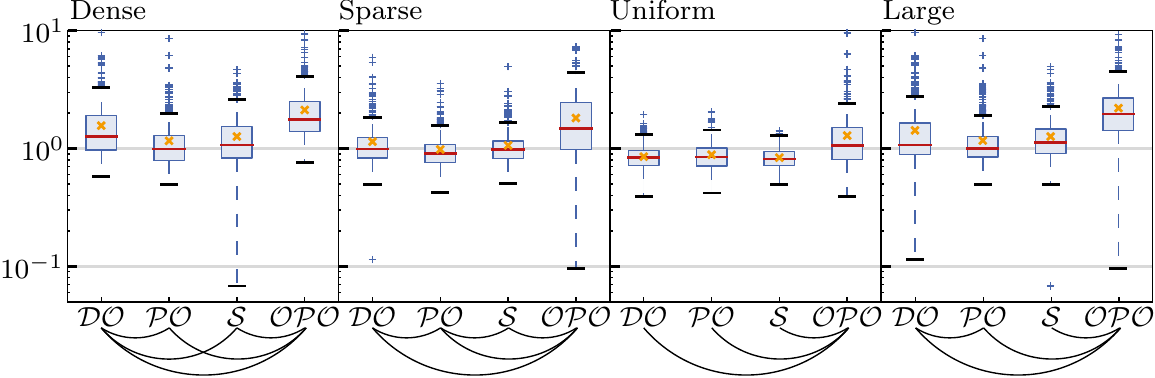}
       \label{fig:response-times}
   }
   \quad
   \subfloat[Success rates. Higher values are better than smaller values.] {
       \includegraphics[page=2,scale=0.95]{fig/plots}
       \label{fig:accuracy}
   }
   \caption{Performance results broken down to dense, sparse and
     uniform sets as well as to large instances (30 sites). Mean values
     are indicated by 
     'x'. Arcs at the bottom show significant differences that were
     found ($p=0.05$).}
 \label{fig:plots}
\end{figure}

\paragraph{Response Times.} Figure~\ref{fig:response-times} shows the
normalized response times broken down into the three considered
distributions $\DB$, $\DC$ and $\DA$, which yield \emph{dense},
\emph{sparse} and \emph{uniform} sets of sites; the corresponding
mean times are found in Table~\ref{table:response-times}, and further plots for both normalized and absolute response times are found in Fig.~\ref{fig:additional:response-times} and Fig.~\ref{fig:additional:abs-response-times} in the appendix.
We obtained
the following results.  Among all leader types, $\LOPO$-leaders have
the highest response time. In particular for dense and sparse sets of
sites the mean response time is up to a factor 1.8 worse than for the
others. For uniform sets we obtain a factor of up to 1.5. Further,
for any distribution the measured differences are
significant. Comparing the response times of the remaining leader
types we obtain the order $\LPO < \LS < \LDO$ with respect to
increasing mean response time. For uniform sets  we did
not measure any pairwise significant difference between~$\LDO$,
$\LPO$ and $\LS$ leaders. However, for dense and sparse sets
we obtained the significant differences as shown in
Fig.~\ref{fig:response-times}. We emphasize that for \LPO- and
\LS-leaders significant differences are measured for sparse, but not
for dense sets of sites. In contrast~\LDO- and \LS-leaders have
significant differences for dense sets, but not for sparse
sets. Further, \LPO- and \LDO-leaders have significant
differences in both dense and sparse sets. Altogether, this justifies the ranking $\LPO < \LS < \LDO$ w.r.t.\ increasing mean response time.

Comparing the instances in terms of \TASKA and \TASKB,
the mean response time of \TASKB is slightly lower than that
of~\TASKA.  Filtering out incorrectly processed tasks does not change
the mean response time much and similar results are obtained; see
Table~\ref{table:response-times}. The mean response times of
\emph{large instances} (any instance with 30 sites and dense, sparse
or uniform distribution) are similar to those of dense sets, and the
mean response times of \emph{small instances} (any instance with 15
sites and dense, sparse or uniform distribution) are similar to those
of uniform sets.

\paragraph{Accuracy.}  We computed for each leader type and each
participant the proportion of instances of that type that the
participant solved correctly; see Fig.~\ref{fig:accuracy} and
Table~\ref{table:accuracy}. Further plots are found in
Fig.~\ref{fig:additional:accuracy} in the appendix.  For dense and
sparse sets of sites we observe that \GOPO has success rates around
$86\%$, while the other groups have success rates greater than
$93\%$. In particular the differences between success rates of
\LOPO-leaders and the remaining types are up to $11\%$ and $13\%$ for
dense and sparse sets, respectively. Any of these differences is
significant, while between \GPO, \GDO and \GS no significant accuracy
differences were measured. For uniform sets of sites, on the other
hand, no significant differences were measured and any group has a
success rate greater than $95\%$. Hence, it appears that uniform sets
of sites produce easily readable labelings with any leader type --
unlike dense and sparse instances.

Considering large and small instances
separately, the group \GOPO has a decreased success rate ($81\%$),
while the other groups remain almost unchanged ($>93\%$), which yields
for \GPO and \GOPO a difference of $16\%$. For small instances no
significant differences were measured.  Comparing the instances by
tasks \TASKA and \TASKB, the success rate of \TASKA is slightly better
than that of~\TASKB except for \GOPO. For the mean response times the
contrary is observed.

\subsection{Preference Data}

Table~\ref{tab:personal-opinion:average-grades} shows the average
grades given by the participants
 with respect to the three
questions Q1--Q3.
 Concerning the general aesthetic appeal (question Q1) leaders of type~$\LDO$ received the best grades (1.8), followed by $\LPO$-leaders (grade 2.3). The participants did not particularly like the appearance of $\LS$-leaders (grade 3.3) and generally disliked $\LOPO$-leaders (grade 4.6).
Table~\ref{tab:personal-opinion:stat} in the appendix lists the detailed percentages of participants who graded a particular leader type better, 
\begin{wraptable}[12]{r}{6.5cm}
\caption{Average grades given by the participants with respect to questions Q1--Q3. Smaller values are better than higher values.}
\centering
\begin{tabular}{lcccc}\toprule
         &\quad $\LDO$\quad &\quad $\LOPO$\quad &\quad $\LPO$ \quad &\quad $\LS$ \\\midrule
{\bf Q1} &\quad 1.8    & \quad 4.6     & \quad 2.3    & \quad 3.3   \\
{\bf Q2} &\quad 2.0    & \quad 4.6     & \quad 2.1    & \quad 2.4   \\
{\bf Q3} &\quad 1.7    & \quad 4.3     & \quad 2.3    & \quad 2.4  \\\bottomrule
\end{tabular}
\label{tab:personal-opinion:average-grades}
\end{wraptable}
equally, or worse than another type.
 In addition to the general impression from the average grades it is worth mentioning that
between the two most preferred leader types $\LDO$ and $\LPO$ 48.4\%
preferred $\LDO$ over $\LPO$ and 38.7\% gave the same grades to both
leader types. Compared to the $\LS$-leaders, a great majority ($>$
80\%) strictly prefers both $\LDO$- and $\LPO$-leaders. 
 In the
interviews seven out of eight participants stated that $\LOPO$-leaders are
``confusing, because leaders closely pass by each other''. They
disliked the long parallel segments of $\LOPO$-leaders. Further, some
participants remarked that \LOPO-leaders ``consist of too many
bends''. For six participants \LS-leaders were ``chaotic and
unstructured'', unlike $\LDO$- and $\LPO$-leaders.  Five participants
said that they liked the flat bend of \LDO-leaders more than the
sharp bend of \LPO-leaders. One participant stated that
``$\LPO$-leaders seem to be more \emph{abstract}
than~$\LDO$-leaders''. Further, it was said that ``the ratio of the
segments' lengths is less balanced for \LPO- than \LDO-leaders.''



For question Q2 (site-to-label) $\LDO$- and $\LPO$-leaders were ranked best (see Table~\ref{tab:personal-opinion:average-grades}), followed by $\LS$ and more than two grades behind by $\LOPO$, whereas for question Q3 (label-to-site) $\LDO$-leaders are further ahead of $\LPO$- and $\LS$-leaders, both of which received similar grades, and are again about two grades ahead of $\LOPO$-leaders.
For questions Q2 and Q3 the most striking observation is that type-$\LS$ leaders received much better results (almost a full grade point better) than for~Q1. This is in strong contrast to the other three leader types, which received grades in the same range as for Q1. This indicates that the participants perceived straight leaders as being well readable during the experiment, but still did not produce very appealing labelings.
%
%
In the interviews participants stated that ``$\LOPO$-leaders are hard
to read because of leaders lying close to each other.'' They
negatively observed that \LOPO-leaders ``may not be clearly
distinguished'', but assessed the ``simple shape of~$\LS$-leaders to
be easily legible.'' Further, they positively noted that ``the
distances between \LDO-leaders seem to be greater than for other
types'' and that ``\LPO-leaders are easier to follow than other
types''.

It is remarkable that the participants rated \LDO-leaders best, while
they ranked third in our performance test. We conjecture
that the participants overestimate the performance of \LDO-leaders, because they
like their aesthetics. For \LS-leaders the reverse is true. In
contrast, their assessment on \LPO- and \LOPO-leaders corresponds more
closely with the result of our performance test.

In summary, $\LDO$-leaders obtained the best subjective ratings. The regularly shaped $\LPO$- and $\LDO$-leaders both scored better than the irregular and less restricted $\LS$-leaders. For any of the three questions
$\LOPO$-leaders were rated a lot worse than the others, which is, according to the interviews, mostly due to the frequent occurrence of many nearby leaders running closely together.

%



\section{Discussion}\label{sec:discussion}

In Section~\ref{sec:hypo} we hypothesized that labelings with many parallel leaders lying close to each other have a significant negative effect on response times and accuracy. Our results from Section~\ref{sec:analysis-user-studie} indeed support hypotheses (H1b) and (H2), which said that the assignment task has a significantly smaller response time and error rate for $\LDO$- and $\LPO$-labelings than for $\LOPO$-labelings in dense (H1b) and also sparse sets of sites (H2). Hypothesis (H2) was claimed to also hold for $\LS$-labelings versus $\LOPO$-labelings, which is confirmed by the experiment as well.
While greater
response times may still be acceptable in some cases, the significantly lower accuracy clearly restricts the usability of $\LOPO$-leaders. Only for small numbers of sites and uniform distributions~$\LOPO$-leaders have comparable
success rates to the other leader types. This judgment is strengthened further by the preference ratings. On average the participants graded $\LOPO$-leaders between 4 (sufficient) and 5 (poor) in all concerns.
The main reason given in the interviews was that $\LOPO$-labelings are confusing due to many leaders closely passing by each other.

However, our results falsified hypothesis (H1a), which claimed that for dense instances type-$\LS$ leaders perform significantly better than the other three leader types. Rather we gained unexpected
insights into the readability of boundary labeling. While we had
expected that due to their simple shape and easily distinguishable slopes \LS-leaders will perform
better than all other types of leaders, we could not measure
significant differences between \LPO-leaders and
\LS-leaders. Interestingly, on average, the participants graded
\LPO-leaders better than \LS-leaders in all examined concerns, in
particular with respect to their aesthetics (Q1). This is emphasized
by the statements given by the participants that $\LPO$-labelings
appear structured while \LS-labelings were perceived as chaotic.
Comparing \LDO- and \LS-leaders we measured some evidence for
(H1a), namely that the assignment task has significantly smaller
response times for \LS- than for \LDO-leaders. However, the
success rates did not differ significantly.

We summarize our main findings regarding the four leader types as follows:
\begin{compactenum}[(1)]
\item \LDO-leaders perform best in the preference rankings, but concerning
  the assignment tasks they perform slightly worse than \LPO- and \LS-leaders.
\item \LOPO-leaders perform worst, both in the assignment tasks and the preference rankings. They are applicable only for small instances or for uniformly distributed sites.
\item \LPO-leaders perform best in the assignment tasks, and received
  good grades in the preference rankings.
\item \LS-leaders perform well in the assignment tasks, but not in the preference rankings. The
  participants dislike their unstructured appearance.
\end{compactenum}
%

We can generally recommend $\LPO$-leaders as the best compromise between measured task performance and subjective preference ratings. For aesthetic reasons, it may also be advisable to use $\LDO$-leaders instead as they have only slightly lower readability scores but are considered the most appealing leader type.

An interesting question is why type-$\LS$ leaders (which showed good task performance) are frequently used by professional graphic designers, e.g., in anatomical drawings, although they were not perceived as aesthetically pleasing in our experiment. One explanation may be that our experiment judged all leader types on an empty background, where the leaders receive the entire visual attention of a viewer. In reality, the labeled figure itself is the main visual element and the leaders should be as unobtrusive as possible and not interfere with the figure. It would be necessary to conduct further experiments to assess the influence and interplay of image and leaders on more complex readability tasks.

Another interesting follow-up question is whether the chosen objective function produces actually the most aesthetic and most readable labelings. Despite being the predominant objective function in the literature on boundary labeling, simply minimizing the total leader length most certainly does not capture all relevant quality criteria.

%
%

\medskip
\noindent\textbf{Acknowledgments.} We thank Helen Purchase and Janet
Siegmund for their advice on the statistical data analysis.

\newpage
\appendix

\section{Data}

\begin{table}[htbp]
  \caption{Mean normalized response times for overall processed tasks and for correctly processed tasks.
 The times are broken down into dense, sparse and uniform sets of sites, into large and small instances, as well as the tasks \TASKA and \TASKB. }
  \centering
  \begin{tabular}{lccccccccccccccccc}
   \toprule
   & \multicolumn{4}{c}{{\it Overall processed tasks}} & \multicolumn{4}{c}{{\it Correctly processed tasks}}\\
   & \GDO & \GPO & \GS & \GOPO &\GDO & \GPO & \GS & \GOPO\\
 \midrule
  Dense & 1.564	&1.161	&1.266	& 2.122 & 1.552 &	1.142&	1.262 & 2.020\\
  Sparse & 1.143 & 0.981 & 1.063&1.813& 1.132 &  0.980 &	1.065 & 1.667\\
  Uniform& 0.852	&0.885	&0.836	&1.287 & 0.855  & 0.894 &	0.837 & 1.231\\
  \midrule
  Large & 1.425 & 1.167 & 1.262 & 2.201 & 1.405 & 1.158 & 1.262 &2.083\\
  Small & 0.949 & 0.852 & 0.848 & 1.281 & 0.948 & 0.854 & 0.850 & 1.239\\
  \midrule
  \TASKA & 1.276 & 1.083 & 1.074 & 1.748 & 1.266 & 1.086 & 1.072 & 1.602 \\
  \TASKB & 1.098 & 0.936 & 1.037 & 1.743 & 1,081 & 0.922 & 1.032 & 1.648\\
  \bottomrule
  \end{tabular}
  \label{table:response-times}
\end{table}

\begin{table}[htbp]
  \caption{Mean success rates. The rates are broken down into dense, sparse and uniform sets of sites, into large and small instances, as well as the tasks \TASKA and \TASKB. }
  \centering
  \begin{tabular}{lccccccccccccccccc}
   \toprule
   & \GDO & \GPO & \GS & \GOPO\\
 \midrule
  Dense & 0.930 &  0.973 &	0.952 & 0.860 \\
  Sparse & 0.957  &	0.987 &	0.978 & 0.855\\
  Uniform& 0.981 & 0.973 &	0.987 & 0.952\\
  \midrule
  Large & 0.943 & 0.973 & 0.945 & 0.814 \\
  Small & 0.970 & 0.982 & 0.988 & 0.964 \\
  \midrule
  \TASKA & 0.959 & 0.991 & 0.982 & 0.886 \\
  \TASKB & 0.953 & 0.964 & 0.962 & 0.892\\
  \bottomrule
  \end{tabular}
  \label{table:accuracy}
\end{table}

\begin{table}[htbp]
  \centering
  \caption{Results of the Dunn-Borferroni test on the
response times of the respective pairs of groups.
The values estimate the likelihood that both respective sample groups
are from the same population. A value $<0.05$ is treated as
statistically significant difference (marked green).
    OPT:  all processed tasks. CPT:  correctly processed tasks only. \TASKA/\TASKB: Restricted to instances of task \TASKA/\TASKB. \textit{Dense}/\textit{Sparse}/\textit{Uniform}: Restricted to instances of distribution dense/sparse/uniform. \textit{Large}/\textit{Small}: Restricted to instances containing 30 and 15 sites, respectively.   }
  \label{table:sig-timing}

\centering
\begin{tabular}{lccccccccccc}
\toprule
\multicolumn{1}{l}{} & \GPO-\GS &\quad& \GS-\GDO &\quad & \GPO-\GDO &\quad& \GDO-\GOPO  &\quad& \GS-\GOPO &\quad& \GPO-\GOPO  \\
\hline
OPT:\TASKA            & \cellcolor{insig-bg} $1.0$  &&   \cellcolor{sig-bg} $<10^{-3}$ &&    \cellcolor{sig-bg} $<10^{-3}$ &&   \cellcolor{sig-bg} $<10^{-3}$ &&   \cellcolor{sig-bg} $<10^{-3}$ &&   \cellcolor{sig-bg} $<10^{-3}$ \\
OPT:\TASKB            &   \cellcolor{sig-bg} $0.010$ && \cellcolor{insig-bg} $0.382$    &&    \cellcolor{sig-bg} $<10^{-3}$ &&   \cellcolor{sig-bg} $<10^{-3}$ &&   \cellcolor{sig-bg} $<10^{-3}$ &&   \cellcolor{sig-bg} $<10^{-3}$ \\
OPT:\textit{Dense}    & \cellcolor{insig-bg} $0.415$ &&   \cellcolor{sig-bg} $<10^{-3}$ &&    \cellcolor{sig-bg} $<10^{-3}$ &&   \cellcolor{sig-bg} $<10^{-3}$ &&   \cellcolor{sig-bg} $<10^{-3}$ &&   \cellcolor{sig-bg} $<10^{-3}$  \\
OPT:\textit{Sparse}   & \cellcolor{sig-bg} $0.001$   && \cellcolor{insig-bg} $1.0$     &&  \cellcolor{sig-bg} $<10^{-3}$    &&   \cellcolor{sig-bg} $<10^{-3}$ &&   \cellcolor{sig-bg} $<10^{-3}$ &&   \cellcolor{sig-bg} $<10^{-3}$ \\
OPT:\textit{Uniform}  & \cellcolor{insig-bg} $0.263$ && \cellcolor{insig-bg} $1.0$     &&  \cellcolor{insig-bg} $0.129$    &&   \cellcolor{sig-bg} $<10^{-3}$ &&   \cellcolor{sig-bg} $<10^{-3}$ &&   \cellcolor{sig-bg} $<10^{-3}$ \\
OPT:\textit{Large}  & \cellcolor{insig-bg} $0.335$   && \cellcolor{insig-bg} $0.098$    &&   \cellcolor{sig-bg} $<10^{-3}$    && \cellcolor{sig-bg} $<10^{-3}$    && \cellcolor{sig-bg} $<10^{-3}$  && \cellcolor{sig-bg} $<10^{-3}$  \\
OPT:\textit{Small}   &  \cellcolor{insig-bg} $1.0$  &&  \cellcolor{sig-bg} $0.001$    && \cellcolor{sig-bg} $<10^{-3}$    &&  \cellcolor{sig-bg} $<10^{-3}$    &&  \cellcolor{sig-bg} $<10^{-3}$    && \cellcolor{sig-bg} $<10^{-3}$   \\
\hline
CPT:\TASKA            & \cellcolor{insig-bg} $1.0$  &&   \cellcolor{sig-bg} $0.001$    &&   \cellcolor{sig-bg} $0.001$    &&   \cellcolor{sig-bg} $<10^{-3}$ &&   \cellcolor{sig-bg} $<10^{-3}$ &&   \cellcolor{sig-bg} $<10^{-3}$ \\
CPT:\TASKB            & \cellcolor{insig-bg} $0.281$ && \cellcolor{insig-bg} $0.174$    &&   \cellcolor{sig-bg} $<10^{-3}$ &&   \cellcolor{sig-bg} $<10^{-3}$ &&   \cellcolor{sig-bg} $<10^{-3}$ &&   \cellcolor{sig-bg} $<10^{-3}$ \\
CPT:\textit{Dense}    & \cellcolor{insig-bg} $1.0$  &&   \cellcolor{sig-bg} $<10^{-3}$ &&   \cellcolor{sig-bg} $<10^{-3}$ &&   \cellcolor{sig-bg} $<10^{-3}$ &&   \cellcolor{sig-bg} $<10^{-3}$ &&   \cellcolor{sig-bg} $<10^{-3}$ \\
CPT:\textit{Sparse}   & \cellcolor{sig-bg} $0.002$   && \cellcolor{insig-bg} $1.0$     && \cellcolor{sig-bg} $0.003$      &&   \cellcolor{sig-bg} $<10^{-3}$ &&   \cellcolor{sig-bg} $<10^{-3}$ &&   \cellcolor{sig-bg} $<10^{-3}$ \\
CPT:\textit{Uniform}  & \cellcolor{insig-bg} $0.125$ && \cellcolor{insig-bg} $1.0$     && \cellcolor{insig-bg} $0.135$    &&   \cellcolor{sig-bg} $<10^{-3}$ &&   \cellcolor{sig-bg} $<10^{-3}$ &&   \cellcolor{sig-bg} $<10^{-3}$ \\
CPT:\textit{Large}  & \cellcolor{insig-bg} $1.0$   && \cellcolor{insig-bg} $0.221$  && \cellcolor{sig-bg} $0.013$      &&   \cellcolor{sig-bg} $<10^{-3}$ &&   \cellcolor{sig-bg} $<10^{-3}$ &&   \cellcolor{sig-bg} $<10^{-3}$ \\
CPT:\textit{Small} & \cellcolor{insig-bg} $1.0$   &&  \cellcolor{sig-bg} $0.001$   && \cellcolor{sig-bg} $<10^{-3}$ &&   \cellcolor{sig-bg} $<10^{-3}$ &&   \cellcolor{sig-bg} $<10^{-3}$ &&   \cellcolor{sig-bg} $<10^{-3}$ \\
\bottomrule
\end{tabular}
\end{table}

\begin{table}[htbp]
  \centering
  \caption{Results of the Dunn-Borferroni test on the
success rates of the respective pairs of groups.
The values estimate the likelihood that both respective sample groups
are from the same population. A value $<0.05$ is treated as
statistically significant difference (marked green).
   \TASKA/\TASKB: Restricted to instances of task \TASKA/\TASKB. \textit{Dense}/\textit{Sparse}/\textit{Uniform}: Restricted to instances of distribution dense/sparse/uniform. \textit{Large}/\textit{Small}: Restricted to instances containing 30 and 15 sites, respectively.
   %
}
  \label{table:sig-correctness}

\centering
\begin{tabular}{lccccccccccc}
\toprule
\multicolumn{1}{l}{} & \GPO-\GS &\quad& \GS-\GDO &\quad & \GPO-\GDO &\quad& \GDO-\GOPO  &\quad& \GS-\GOPO &\quad& \GPO-\GOPO  \\
\hline
\TASKA            & \cellcolor{insig-bg} $1.0$  && \cellcolor{insig-bg} $1.0$  && \cellcolor{insig-bg} $0.460$ &&   \cellcolor{sig-bg} $<10^{-3}$ &&   \cellcolor{sig-bg} $<10^{-3}$ &&   \cellcolor{sig-bg} $<10^{-3}$ \\
\TASKB            & \cellcolor{insig-bg} $1.0$  && \cellcolor{insig-bg} $1.0$  && \cellcolor{insig-bg} $1.0$  &&   \cellcolor{sig-bg} $0.001$    &&   \cellcolor{sig-bg} $<10^{-3}$ &&   \cellcolor{sig-bg} $<10^{-3}$ \\
\textit{Dense}    & \cellcolor{insig-bg} $1.0$  && \cellcolor{insig-bg} $1.0$  && \cellcolor{insig-bg} $0.330$ &&   \cellcolor{sig-bg} $0.019$    &&   \cellcolor{sig-bg} $0.001$    &&   \cellcolor{sig-bg} $<10^{-3}$ \\
\textit{Sparse}   & \cellcolor{insig-bg} $1.0$  && \cellcolor{insig-bg} $1.0$  && \cellcolor{insig-bg} $1.0$  &&   \cellcolor{sig-bg} $0.001$    &&   \cellcolor{sig-bg} $<10^{-3}$ &&   \cellcolor{sig-bg} $<10^{-3}$ \\
\textit{Uniform}   & \cellcolor{insig-bg} $1.0$  && \cellcolor{insig-bg} $1.0$  && \cellcolor{insig-bg} $1.0$  && \cellcolor{insig-bg} $0.262$    &&  \cellcolor{insig-bg} $0.125$  &&   \cellcolor{insig-bg} $1.0$ \\
\textit{Large}    & \cellcolor{insig-bg} $1.0$  && \cellcolor{insig-bg} $1.0$  && \cellcolor{insig-bg} $0.922$ &&   \cellcolor{sig-bg} $<10^{-3}$ &&   \cellcolor{sig-bg} $<10^{-3}$ &&   \cellcolor{sig-bg} $<10^{-3}$ \\
\textit{Small}    & \cellcolor{insig-bg} $1.0$  && \cellcolor{insig-bg} $0.764$ && \cellcolor{insig-bg} $1.0$  && \cellcolor{insig-bg} $1.0$     && \cellcolor{insig-bg} $0.055$    && \cellcolor{insig-bg} $0.262$  \\
\bottomrule
\end{tabular}
\end{table}
\FloatBarrier

\begin{table}[htbp]
  \centering
  \caption{Statistics for questions Q1--Q3. The percentage of participants that graded a leader type better ($<$), equally ($=$) or worse ($>$) than another. Majorities are highlighted in bold.}
\begin{tabular}{lccc|ccc|ccc}
\toprule
  &\ $\LDO{<}\LOPO$ & $\LDO{=}\LOPO$ & $\LDO{>}\LOPO$ & $\LDO{<}\LPO$ &  $\LDO{=}\LPO$ & $\LDO{>}\LPO$ & $\LDO{<}\LS$ & $\LDO{=}\LS$ & $\LDO{>}\LS$ \\
 \midrule
  {\bf Q1} & \textbf{100}          & 0            & 0            & \textbf{48.4}        & 38.7        & 12.9        & \textbf{90.3}       & 3.2        & 6.5        \\
  {\bf Q2} & \textbf{93.5}         & 3.2          & 3.2          & 32.3        & \textbf{41.9}        & 25.8        & \textbf{48.4}       & 19.4       & 32.3       \\
  {\bf Q3} & \textbf{93.5}         & 6.5          & 0            & \textbf{54.8}        & 35.5        & 9.7         & \textbf{48.4}       & 35.5       & 16.1 \\
\toprule
         &\ $\LPO{<}\LOPO$\ &\ $\LPO{=}\LOPO$\ &\ $\LPO{>}\LOPO$\ &\ $\LPO{<}\LS$\ &\ $\LPO{=}\LS$\ &\ $\LPO{>}\LS$\ &\ $\LS{<}\LOPO$\ &\ $\LS{=}\LOPO$\ &\ $\LS{>}\LOPO$ \\\midrule
{\bf Q1} & \textbf{100}          & 0            & 0            & \textbf{80.6}       & 3.2        & 16.1       & \textbf{80.6}        & 6.5         & 12.9        \\
{\bf Q2} & \textbf{96.8}         & 3.2          & 0            & 35.5       & 25.8       & \textbf{38.7}       & \textbf{83.9}         & 9.7         & 6.5        \\
{\bf Q3} & \textbf{93.5}         & 3.2          & 3.2          & \textbf{41.9}       & 16.1       & \textbf{41.9}       & \textbf{93.5}         & 3.2         & 3.2     \\
\bottomrule
\end{tabular}
  \label{tab:personal-opinion:stat}
\end{table}

\begin{figure}[htbp]
  \centering
      \begin{tabular}{cccc}
         \includegraphics[scale=0.32]{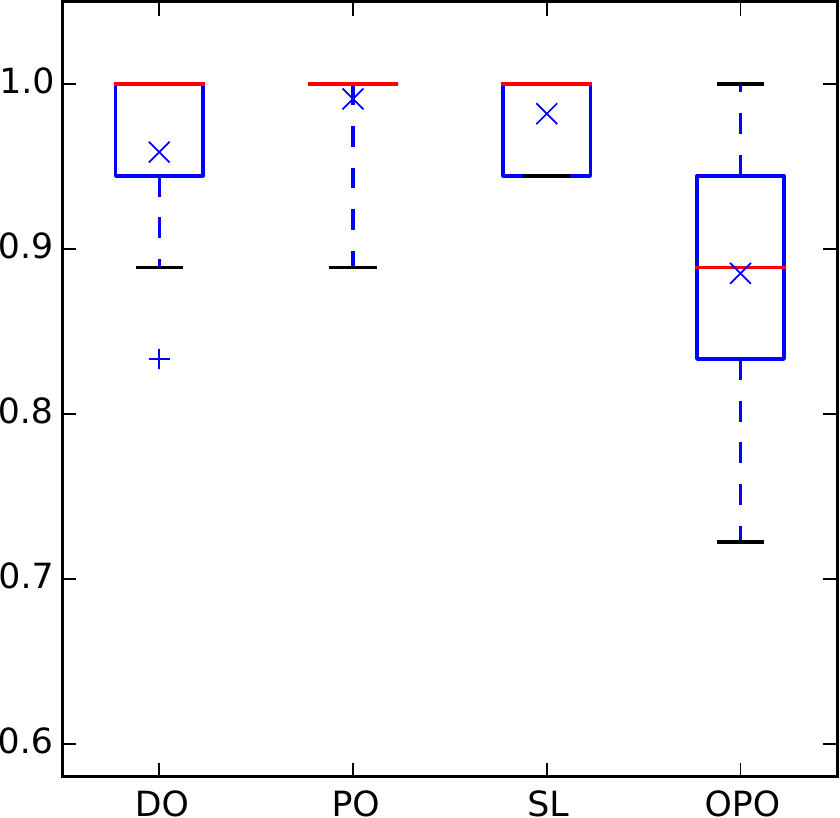}&
         \includegraphics[scale=0.32]{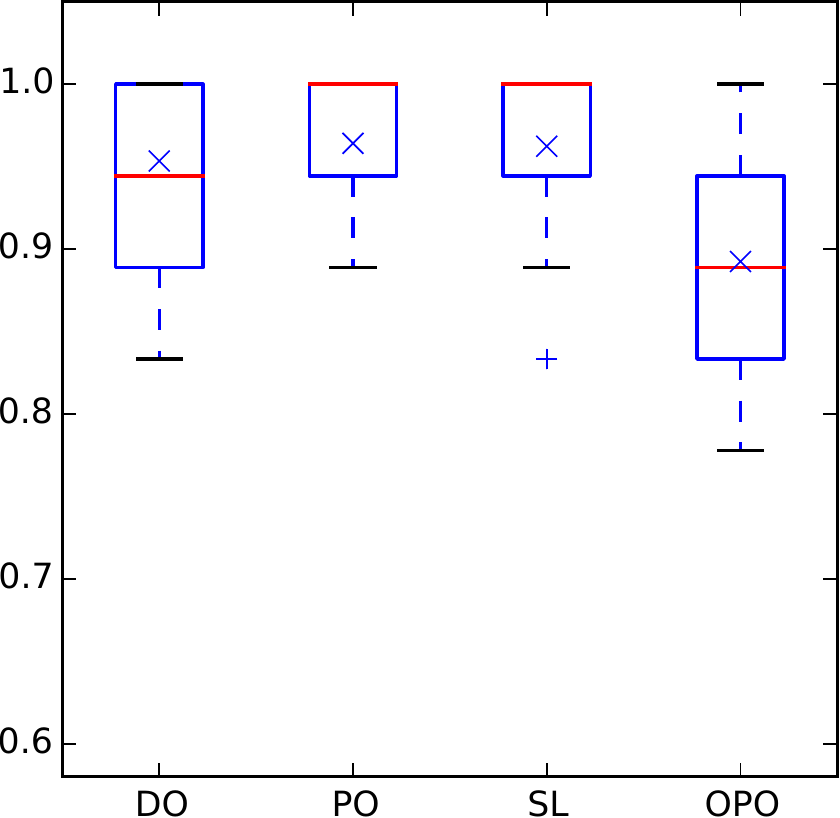}&
         \includegraphics[scale=0.32]{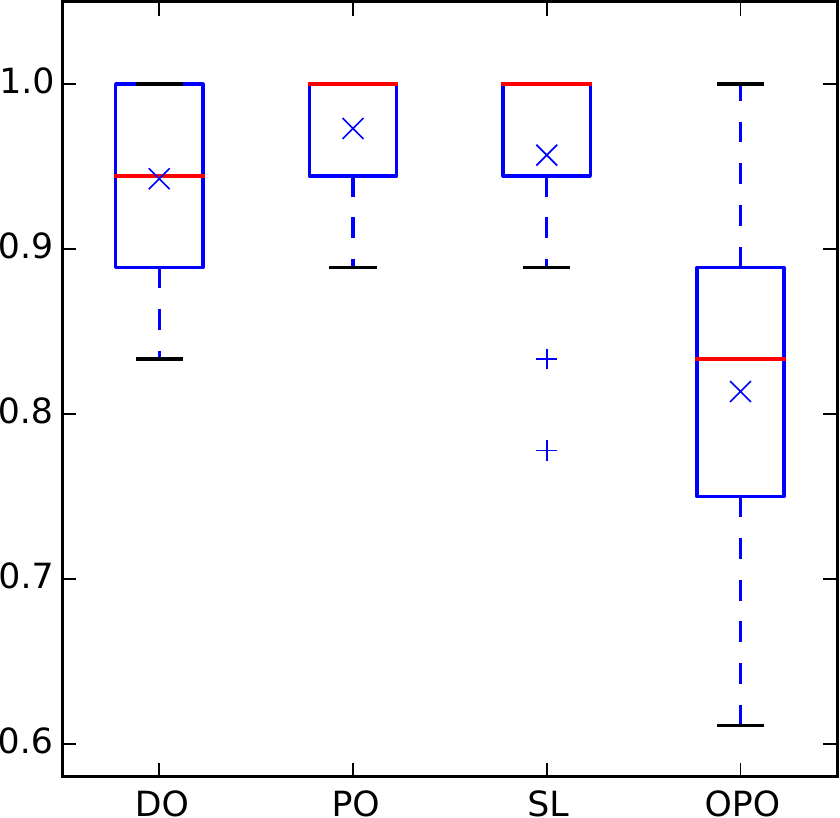}&
         \includegraphics[scale=0.32]{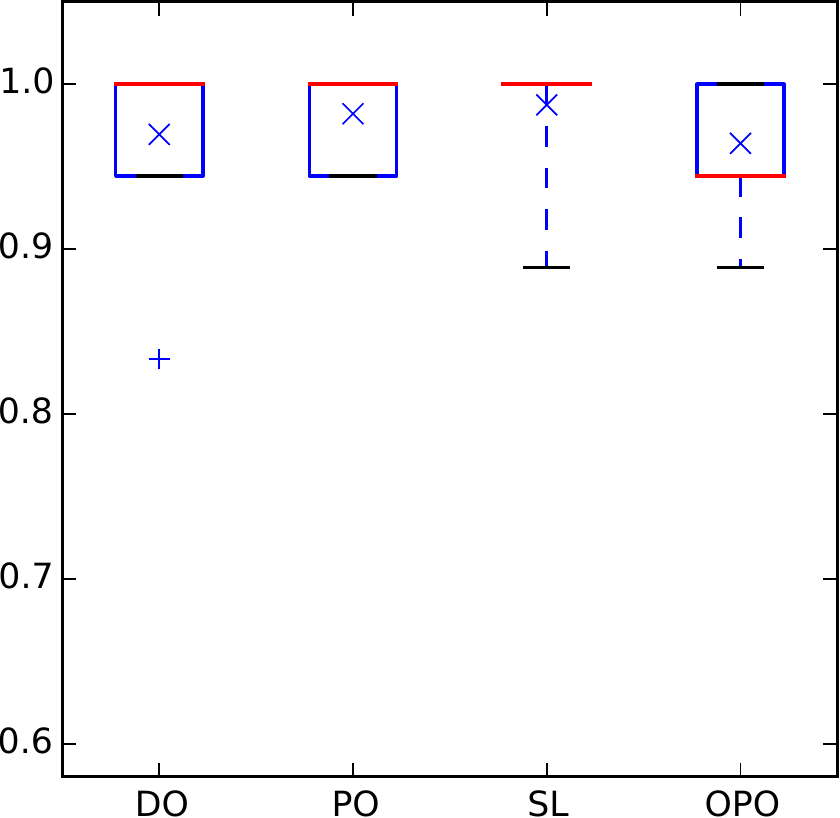}\\
         \TASKA  & \TASKB  & Large & Small
      \end{tabular}
   \caption{ Success rates broken down to different
     parameters. Mean values are indicated by a bold 'x'. The corresponding significances are found in Table~\ref{table:sig-correctness}. Higher values are better than smaller values.}
   \label{fig:additional:accuracy}
\end{figure}

\begin{figure}[htbp]
  \centering
   \subfloat[Response times over all tasks (OPT).]{
      \begin{tabular}{cccc}
         \includegraphics[scale=0.32]{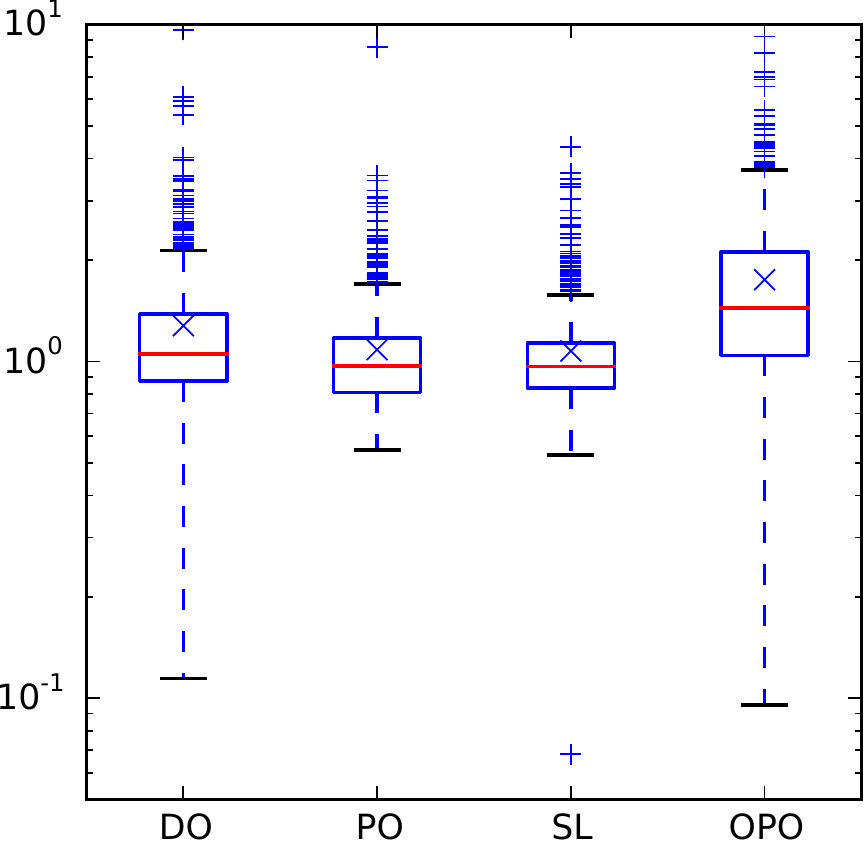}&
         \includegraphics[scale=0.32]{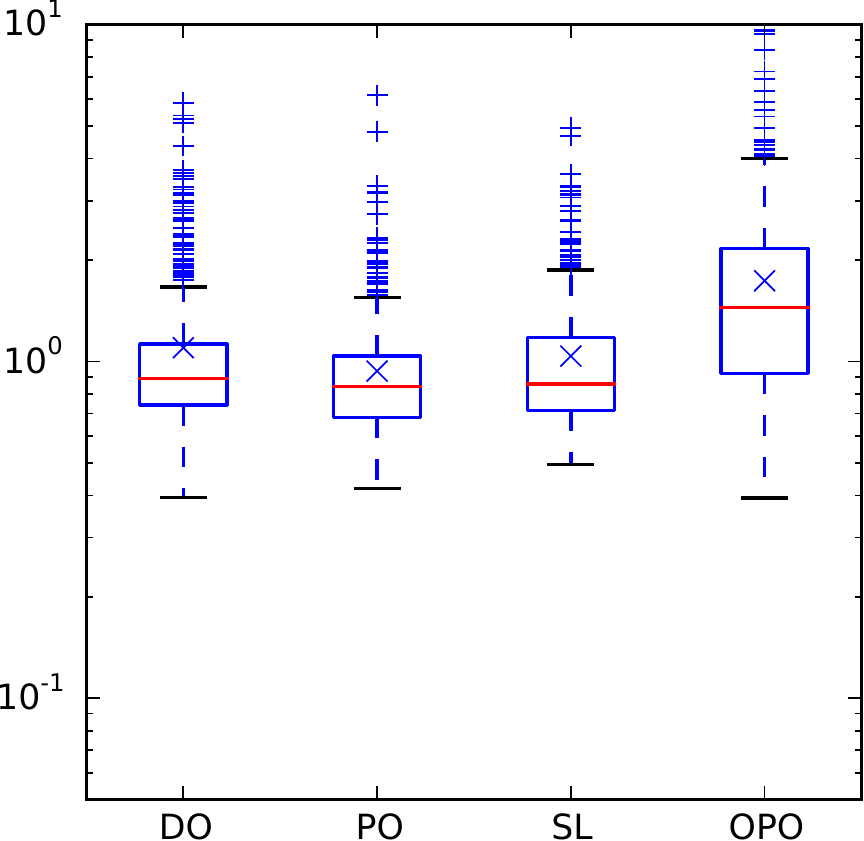}&
         \includegraphics[scale=0.32]{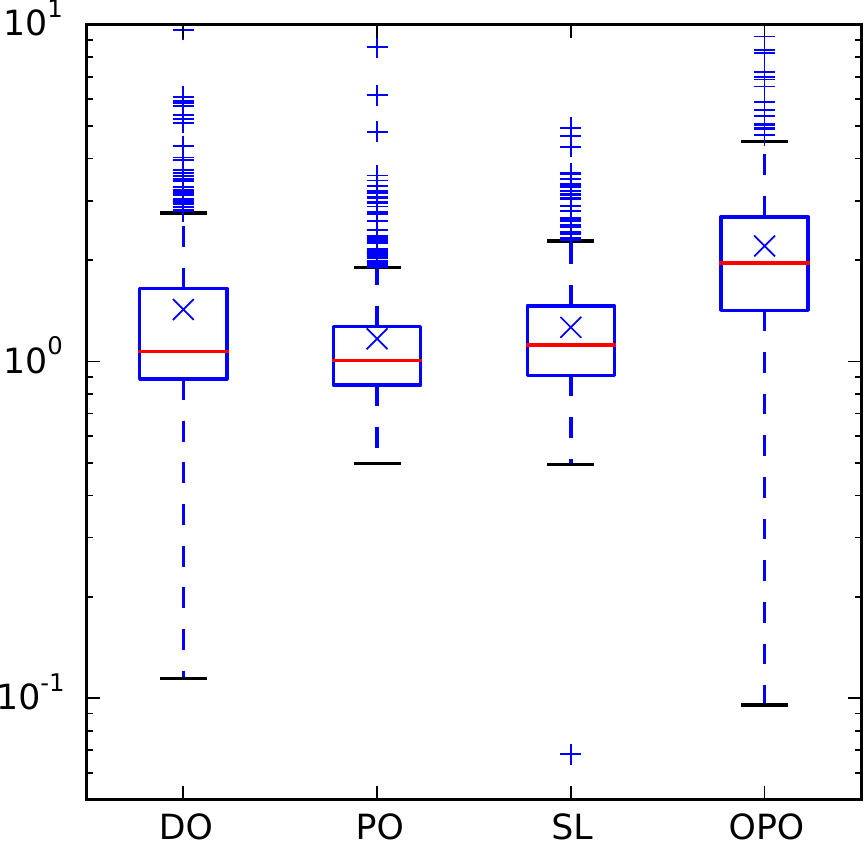}&
         \includegraphics[scale=0.32]{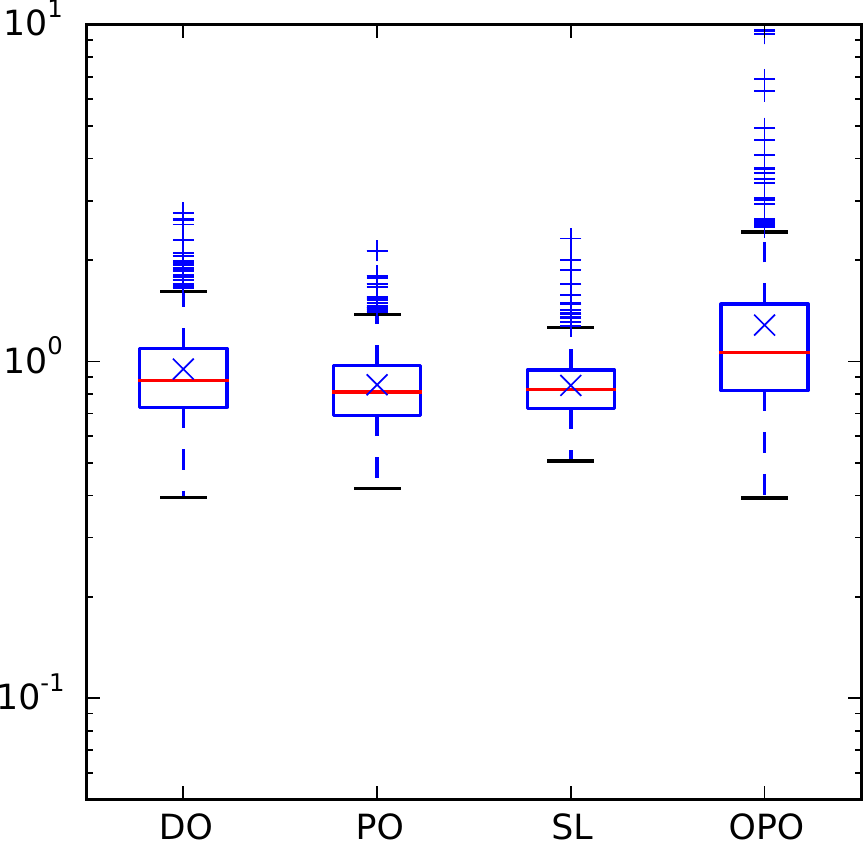}\\
         \TASKA & \TASKB  & Large & Small
      \end{tabular}
   }

   \subfloat[Response times over all correctly processed tasks (CPT).]{
      \begin{tabular}{cccc}
            \includegraphics[scale=0.32]{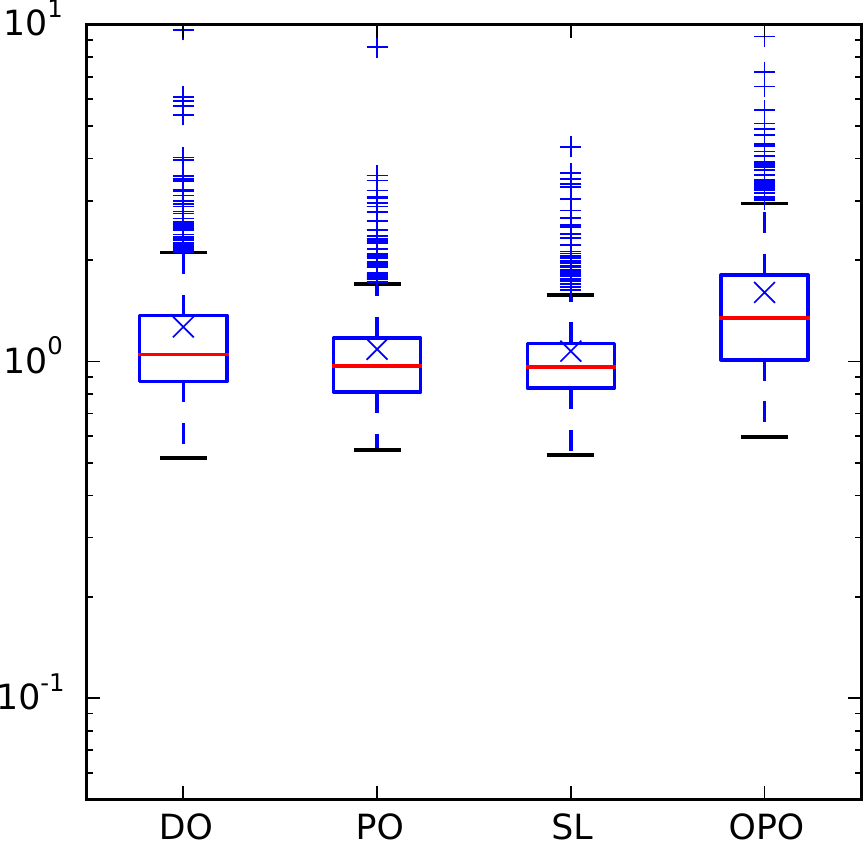}&
            \includegraphics[scale=0.32]{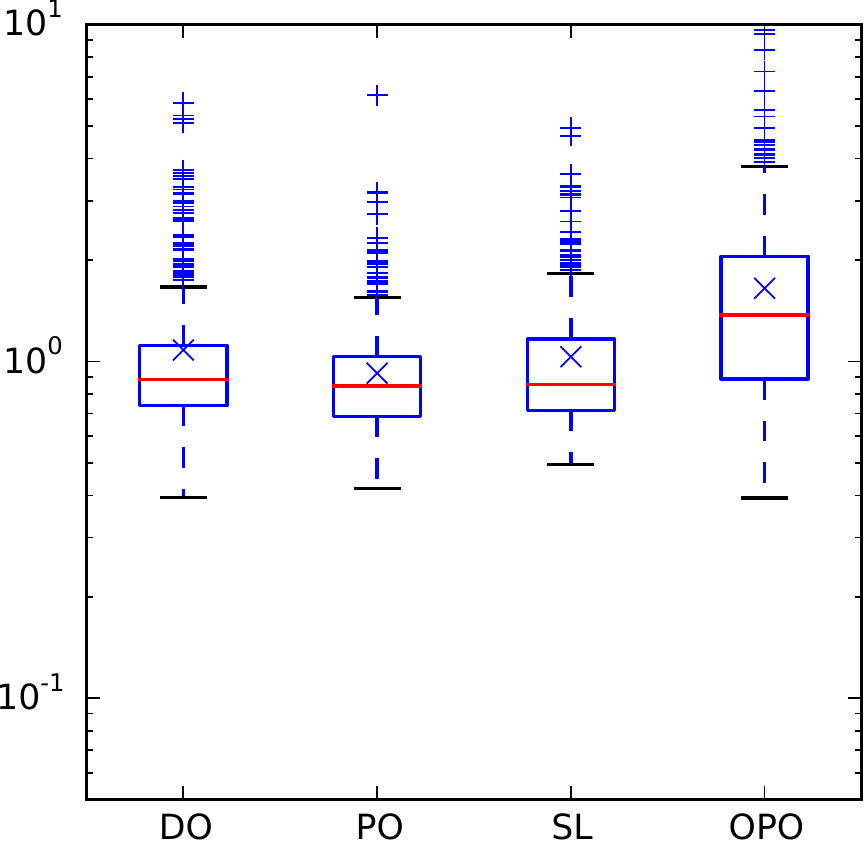}&
         \includegraphics[scale=0.32]{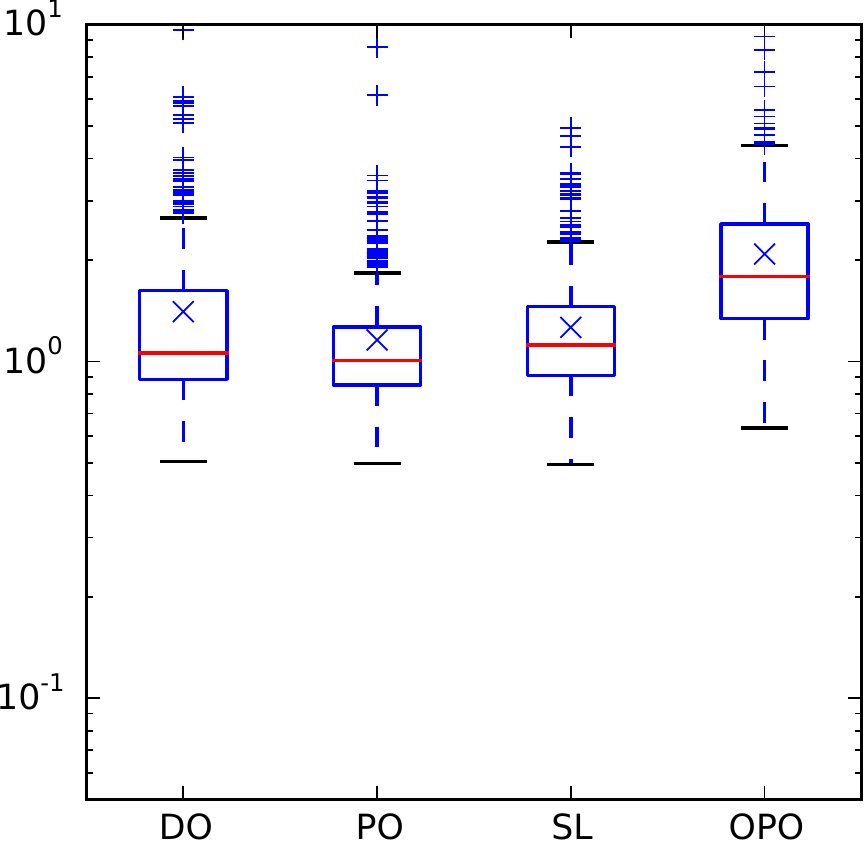}&
         \includegraphics[scale=0.32]{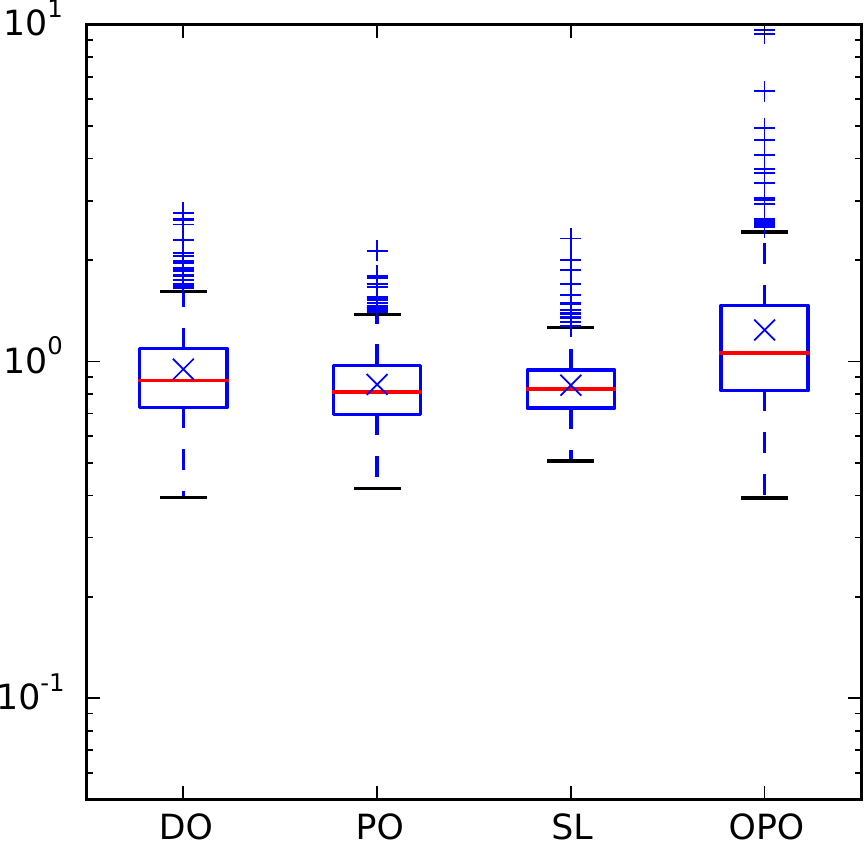}\\
         \TASKA  & \TASKB  & Large & Small
      \end{tabular}
   }

   \subfloat[Response times over all correctly processed tasks (CPT).]{
      \begin{tabular}{cccc}
         \includegraphics[scale=0.32]{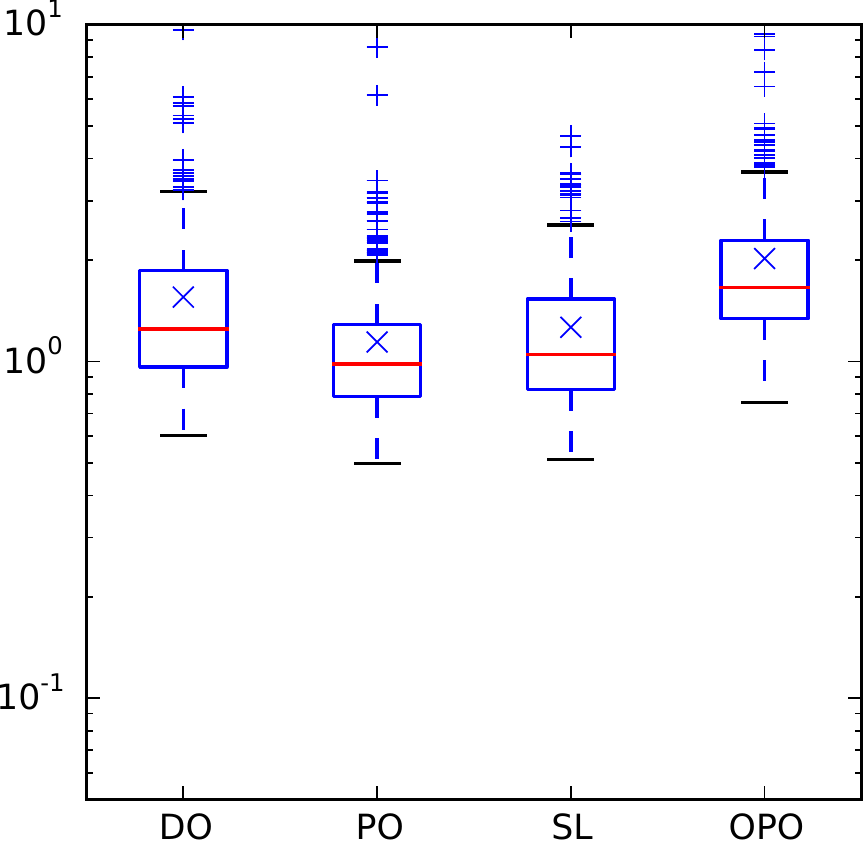}&
         \includegraphics[scale=0.32]{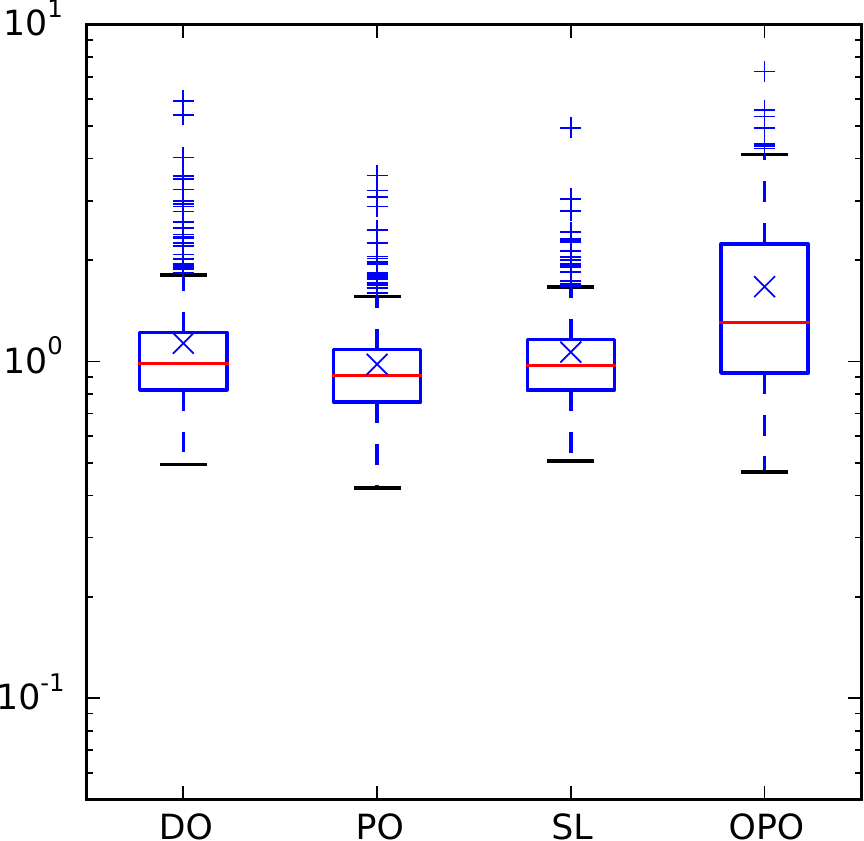}&
         \includegraphics[scale=0.32]{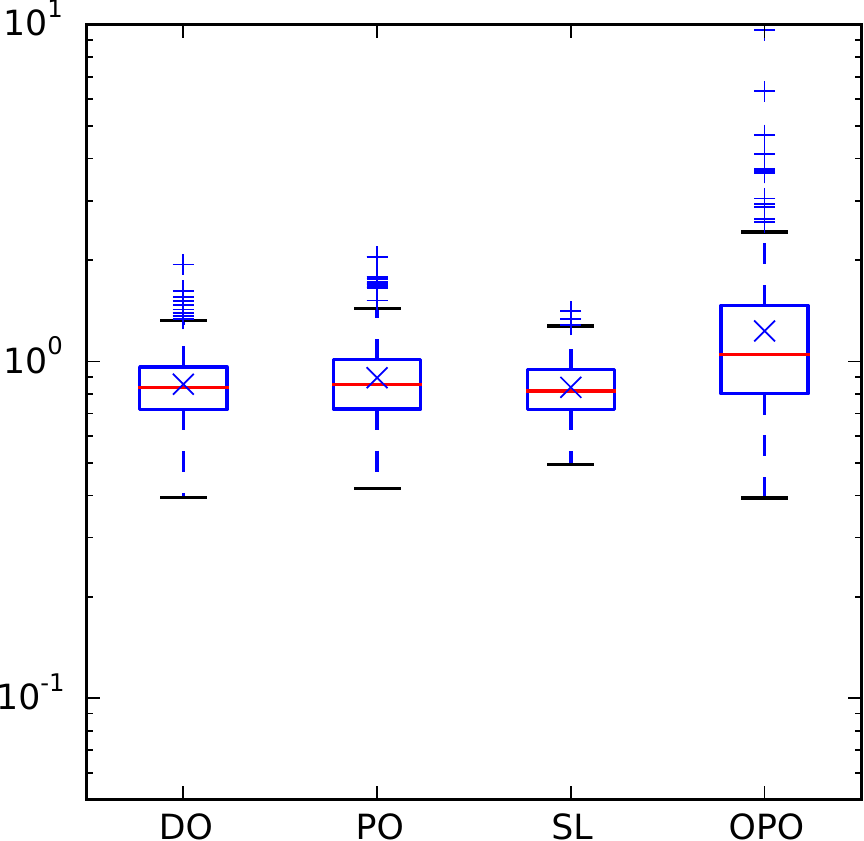}\\
         Dense & Sparse & Uniform
      \end{tabular}
   }
   \caption{ Normalized response times (on log-scale) broken down to different
     parameters. Mean values are indicated by a bold 'x'. The corresponding significances are found in Table~\ref{table:sig-timing}. Smaller values are better than higher values.}
   \label{fig:additional:response-times}
\end{figure}

\begin{figure}[htbp]
  \centering
   \subfloat[Response times in seconds over all tasks (OPT) broken into large and small instances as well as instances for task $\TASKA$ and $\TASKB$.]{
      \begin{tabular}{cccc}
         \includegraphics[scale=0.33]{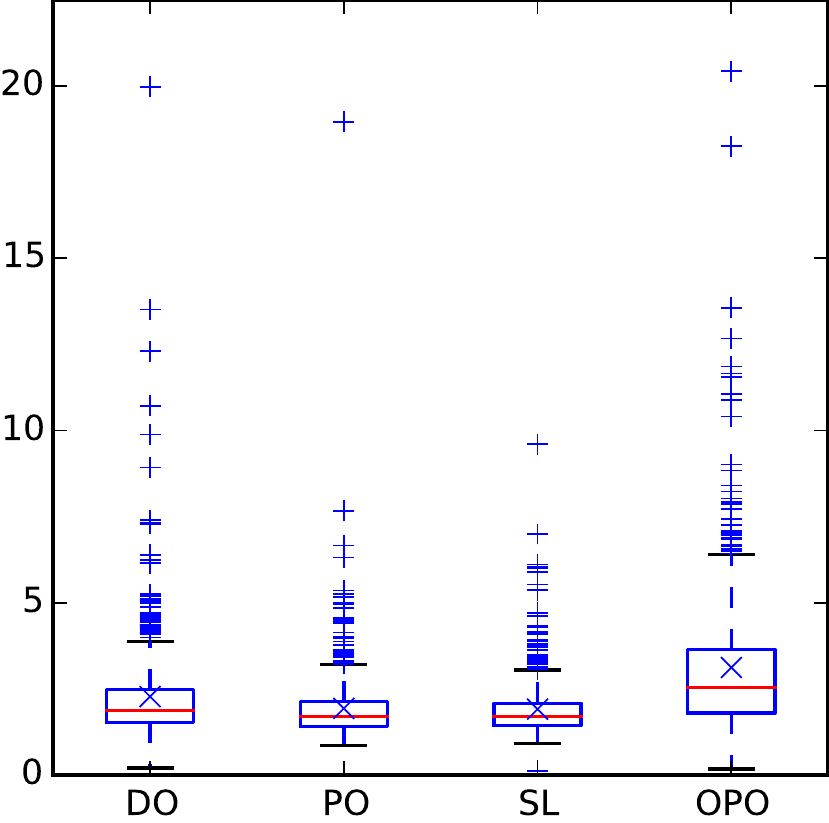}&
         \includegraphics[scale=0.33]{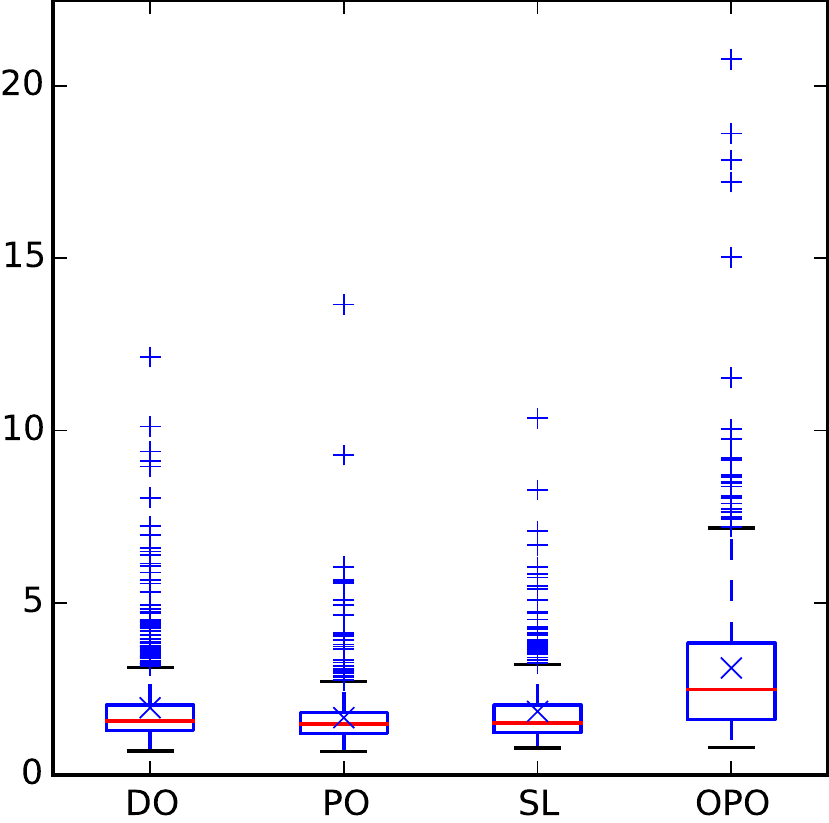}&
         \includegraphics[scale=0.33]{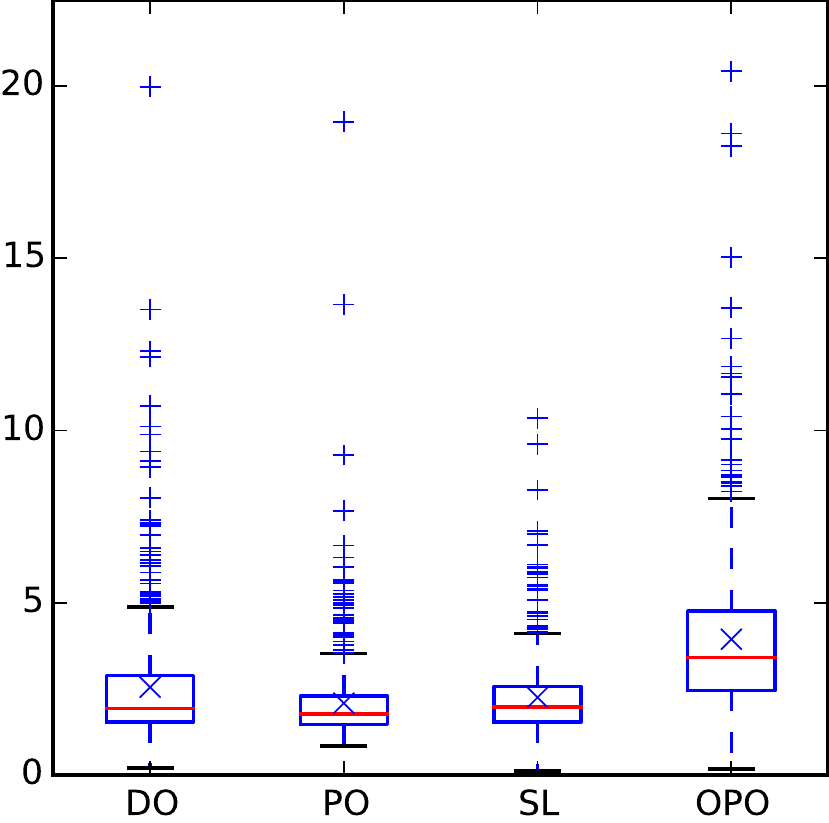}&
         \includegraphics[scale=0.33]{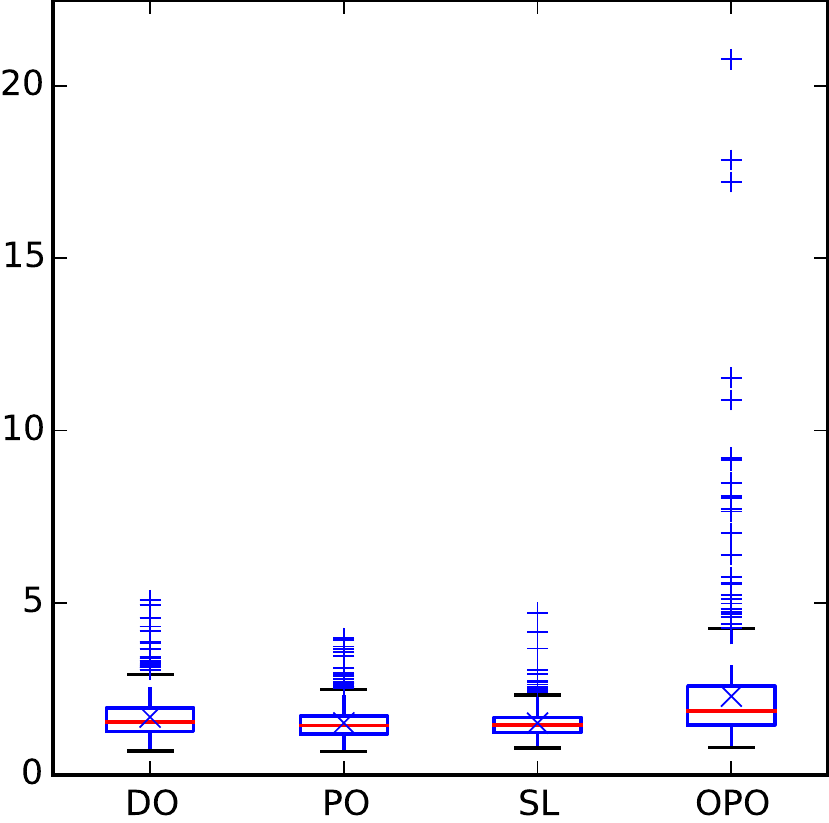}\\
         \TASKA & \TASKB  & Large & Small
      \end{tabular}
   }

  \subfloat[Response times in seconds over all tasks (OPT) broken into dense, sparse and uniform instances as well as all instances.]{
      \begin{tabular}{cccc}
         \includegraphics[scale=0.33]{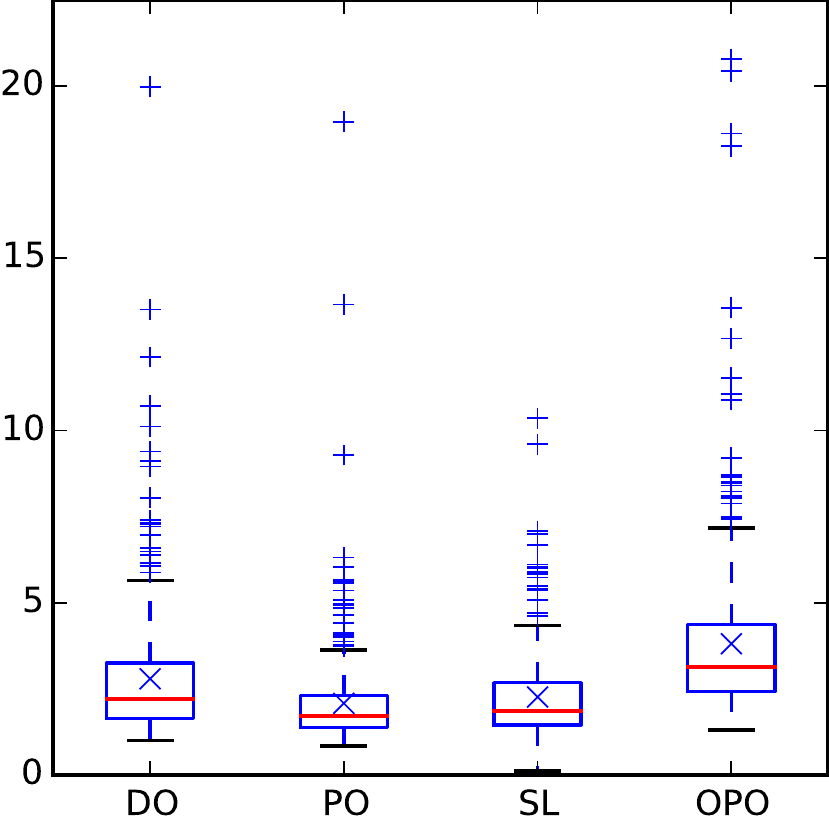}&
         \includegraphics[scale=0.33]{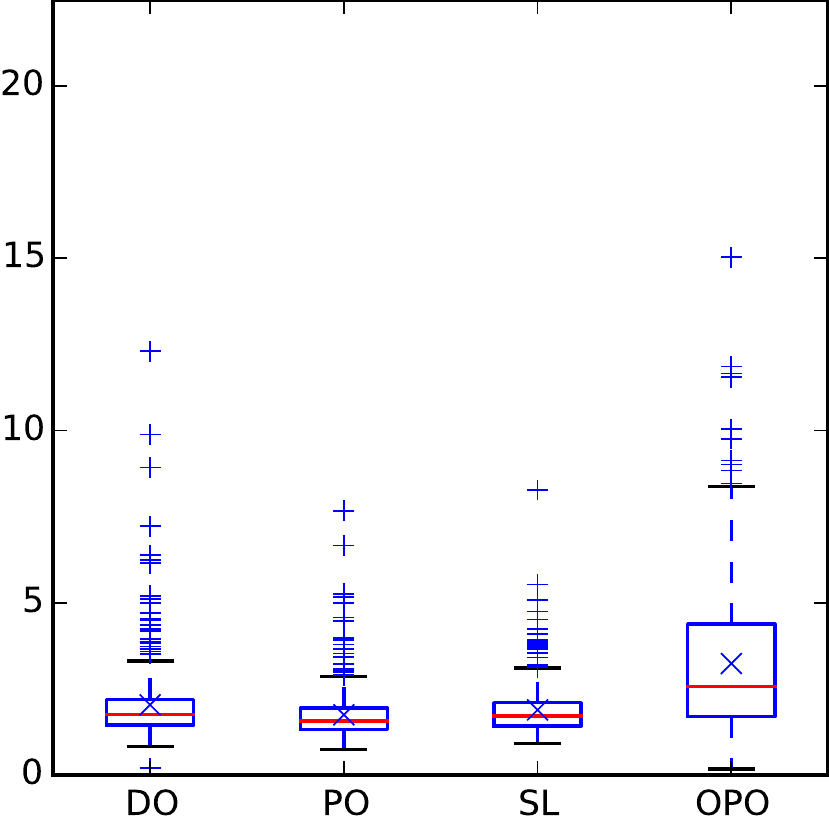}&
         \includegraphics[scale=0.33]{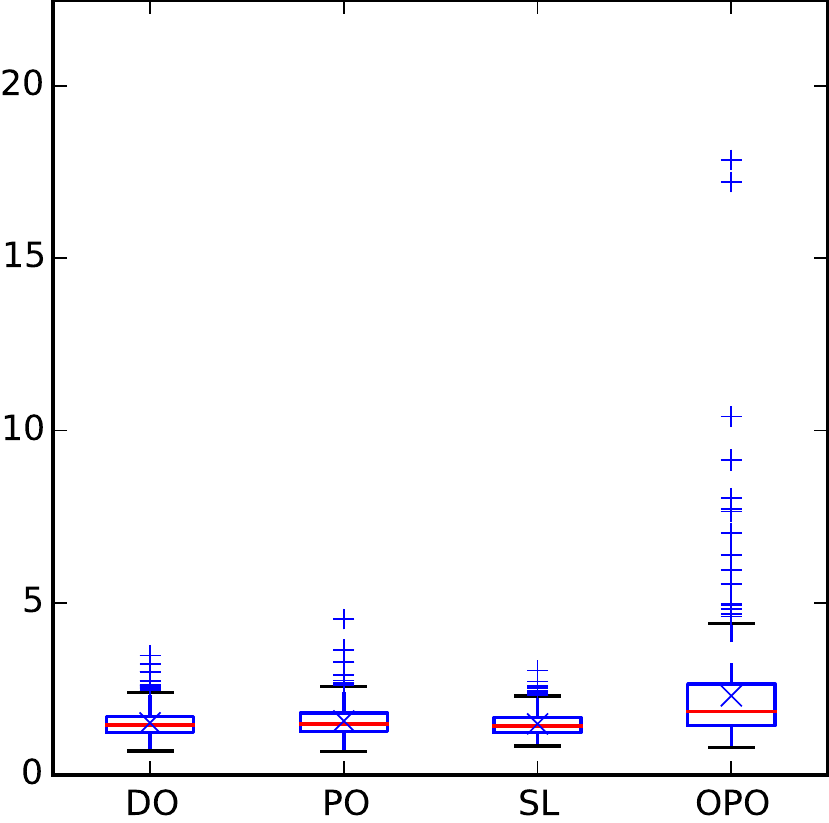}&
         \includegraphics[scale=0.33]{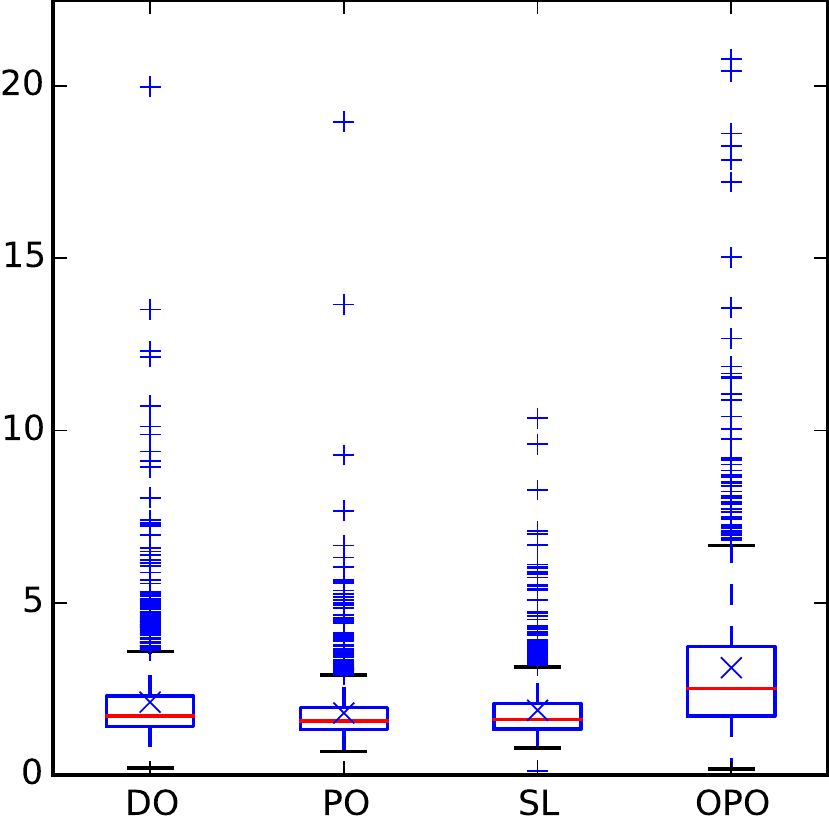}\\
         Dense & Sparse & Uniform & All
      \end{tabular}
   }

   \subfloat[Response times in seconds over all correctly processed tasks (CPT) broken into large, small instances as well as instances for task $\TASKA$ and $\TASKB$.]{
      \begin{tabular}{cccc}
            \includegraphics[scale=0.33]{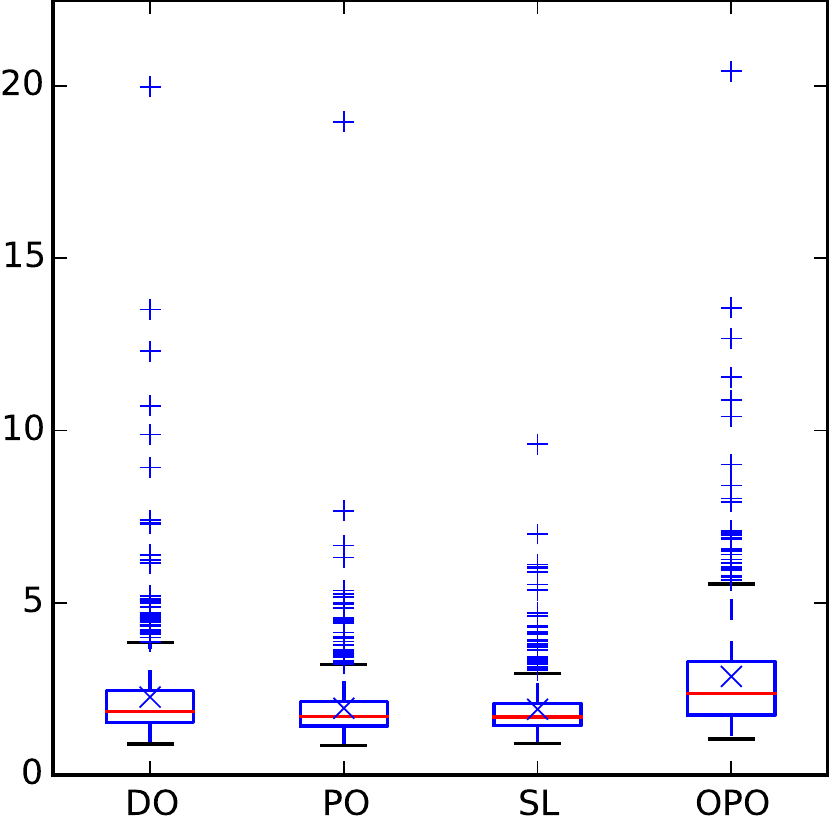}&
            \includegraphics[scale=0.33]{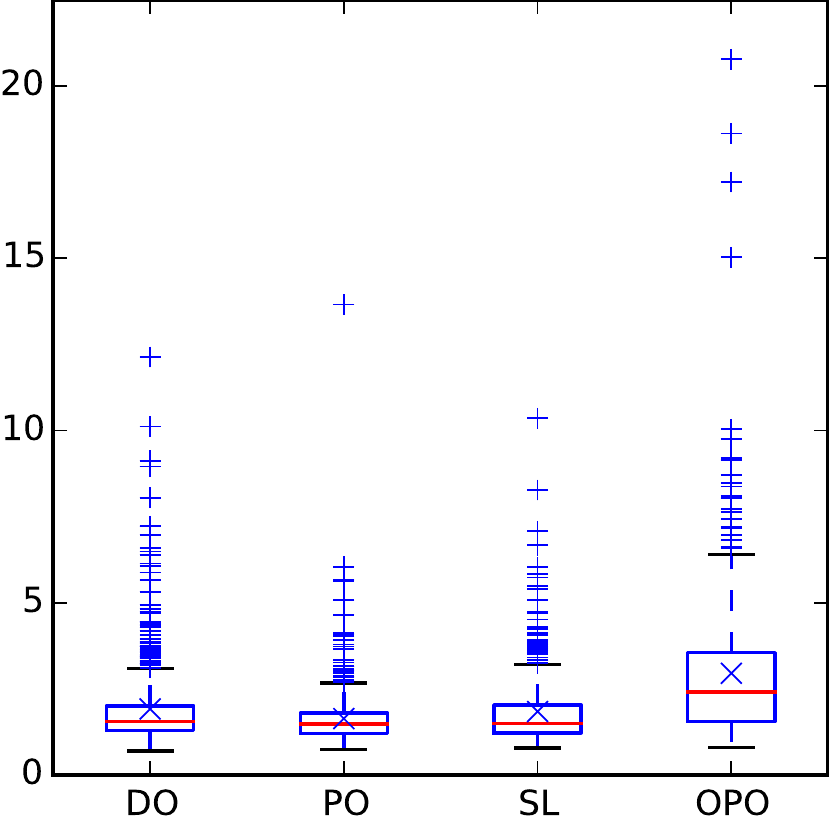}&
         \includegraphics[scale=0.33]{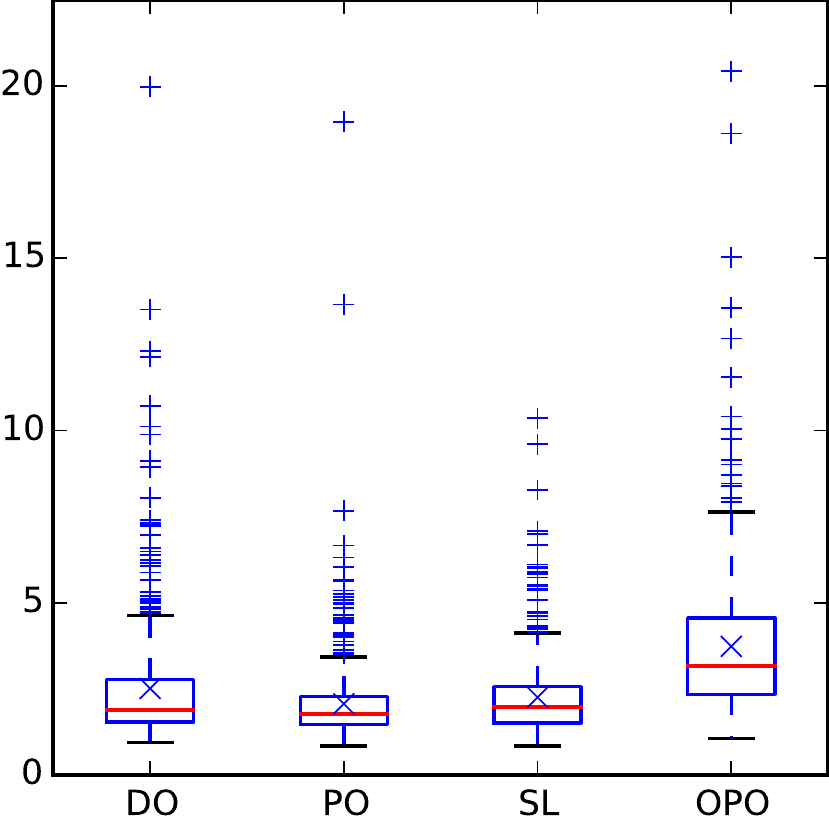}&
         \includegraphics[scale=0.33]{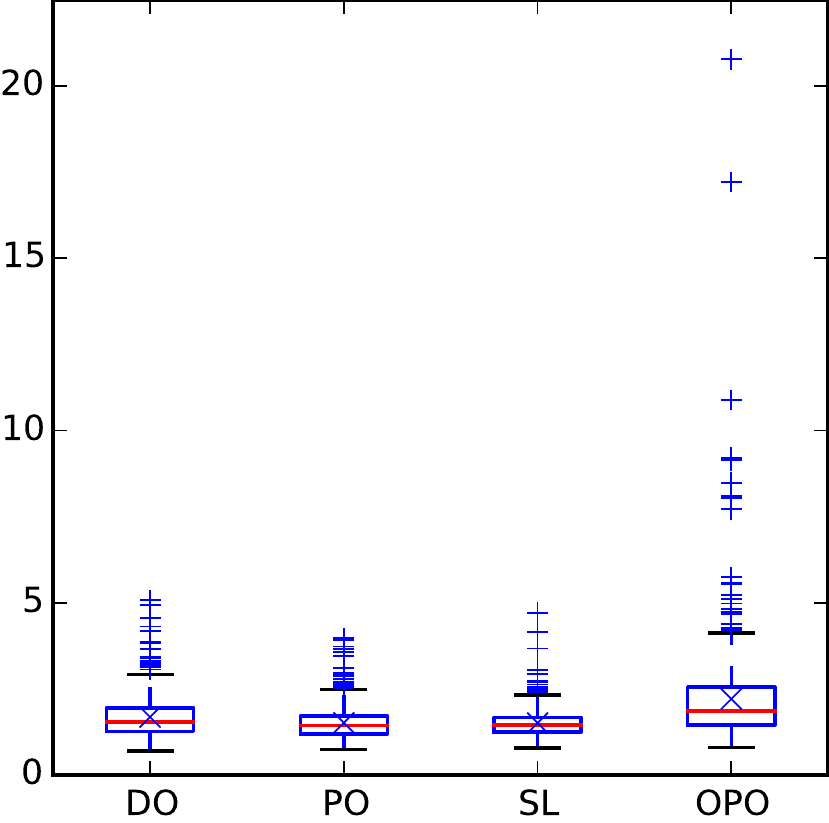}\\
         \TASKA  & \TASKB  & Large & Small
      \end{tabular}
   }

   \subfloat[Response times in seconds over all correctly processed tasks (CPT) broken into dense, sparse and uniform instances as well as all instances.]{
      \begin{tabular}{cccc}
         \includegraphics[scale=0.33]{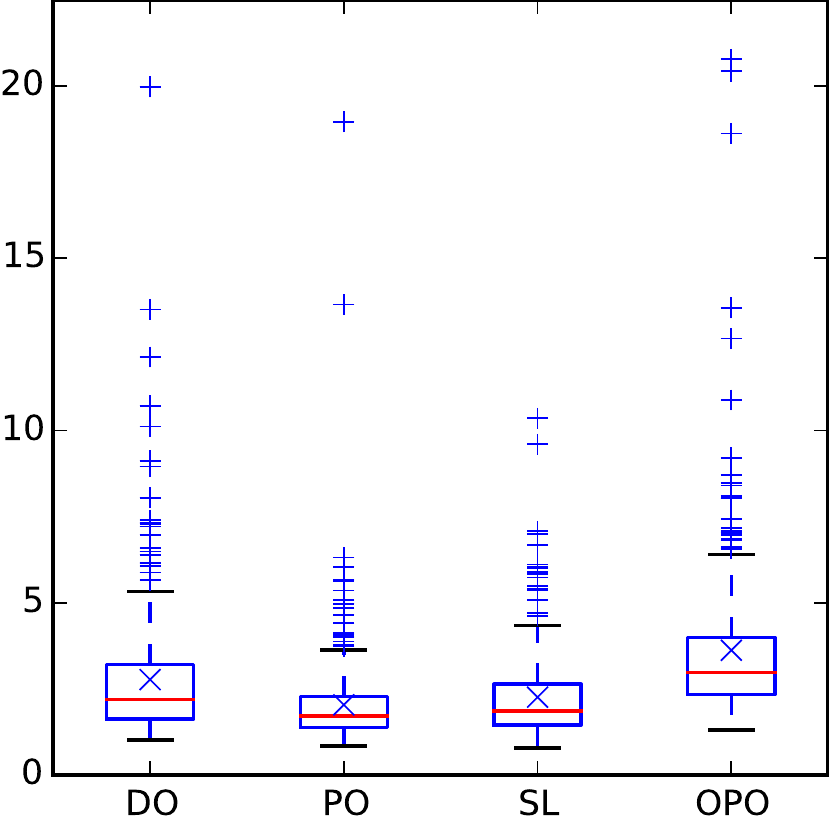}&
         \includegraphics[scale=0.33]{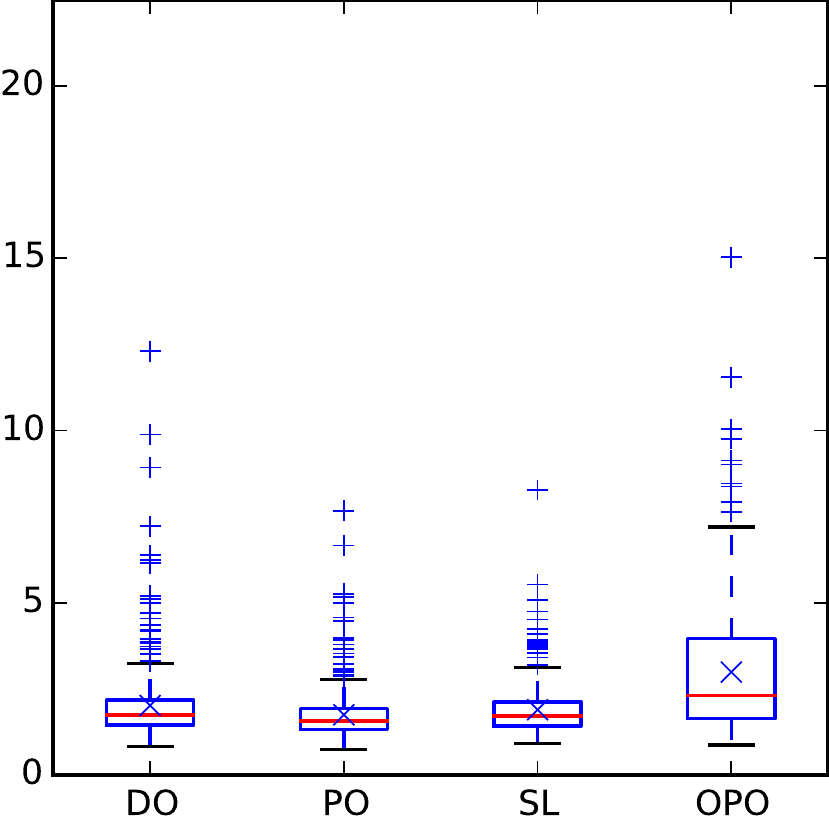}&
         \includegraphics[scale=0.33]{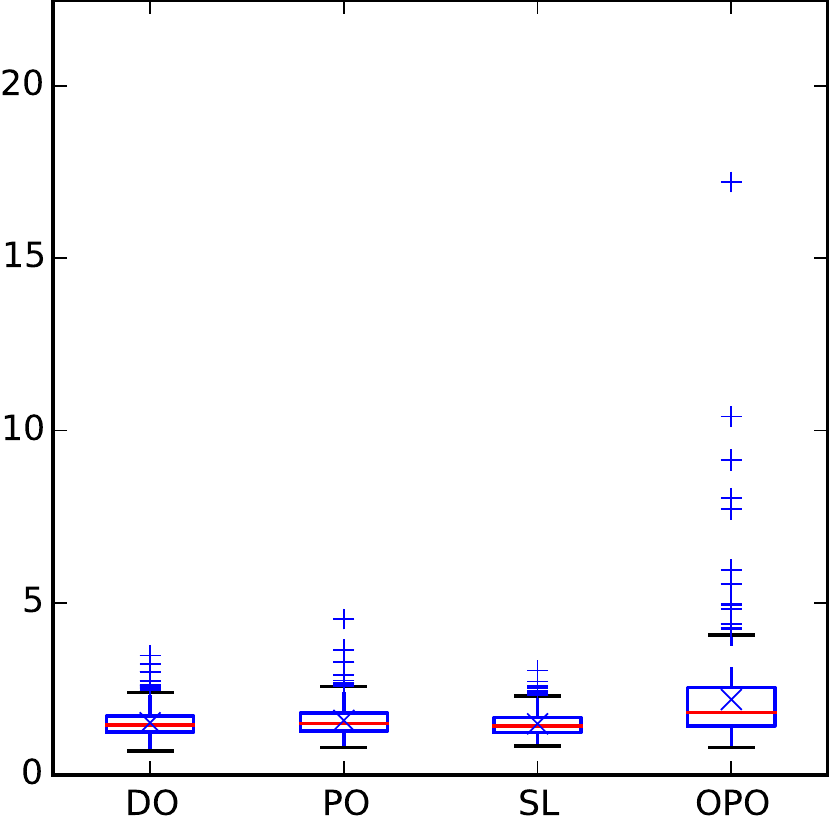}&
         \includegraphics[scale=0.33]{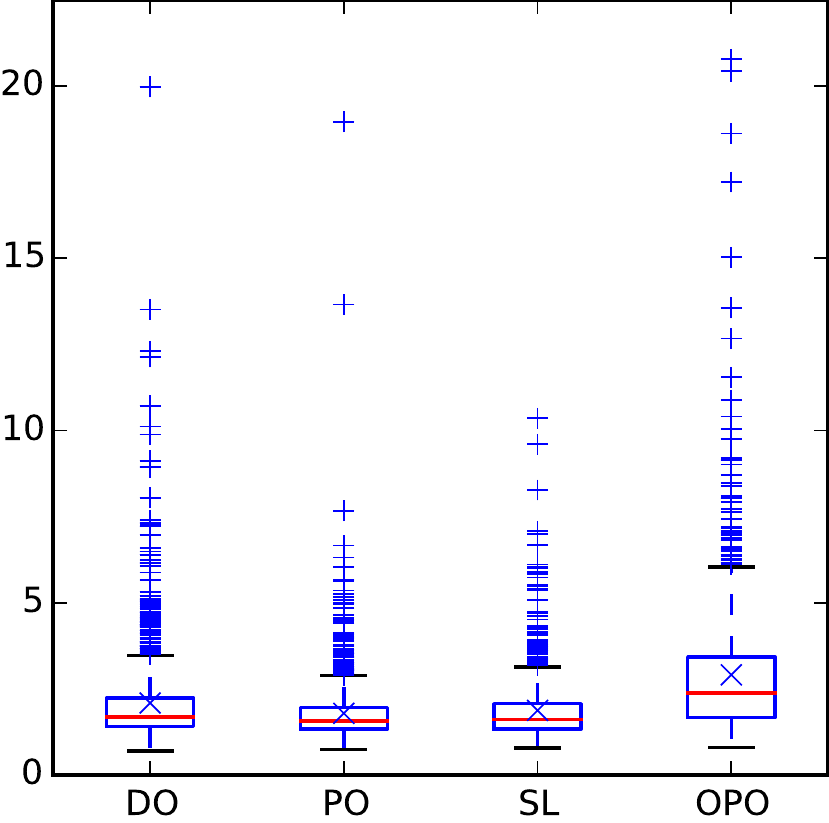}\\
         Dense & Sparse & Uniform & All
      \end{tabular}
   }
   \caption{ Absolute response times (in seconds) broken down to different
     parameters. Mean values are indicated by a bold 'x'. The corresponding significances are found in Table~\ref{table:sig-timing}. Smaller values are better than higher values.}
   \label{fig:additional:abs-response-times}
\end{figure}

\FloatBarrier
\clearpage

\section{Examples of Stimuli}
\begin{figure}[htbp]
\begin{center}
  \begin{tabular}{cc}
    \includegraphics[width=0.3\linewidth]{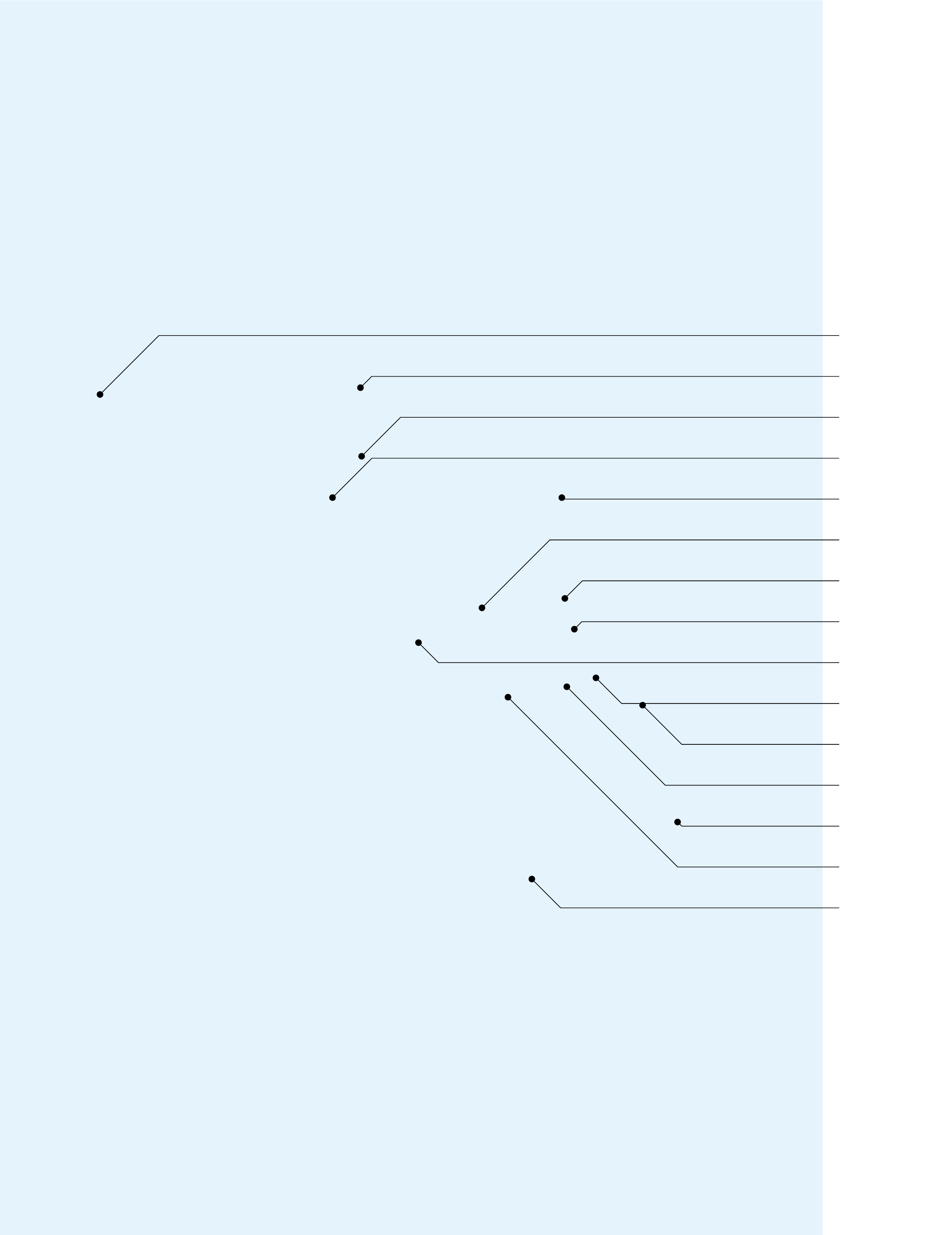} & \includegraphics[width=0.3\linewidth]{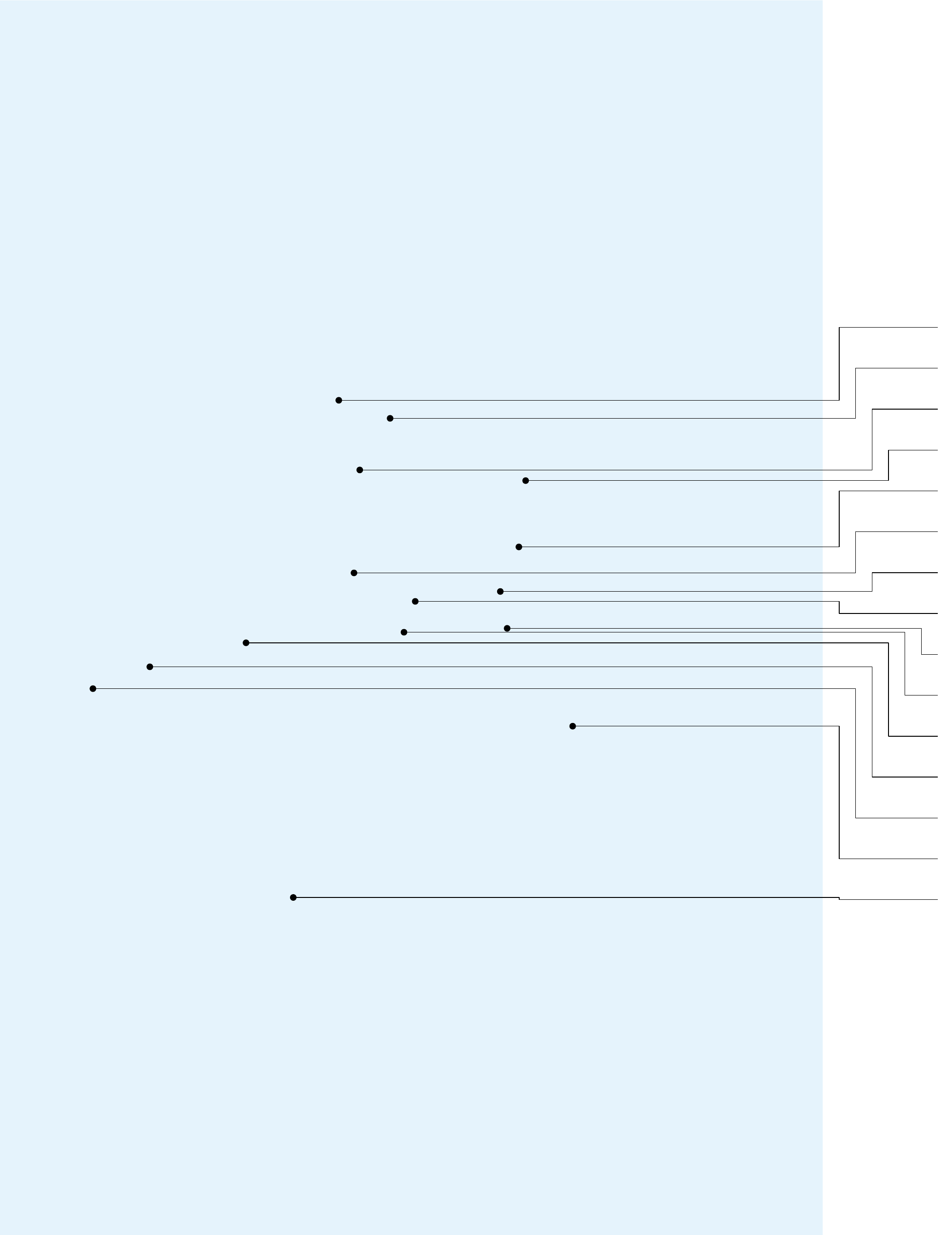}\\
    \LDO-leaders & \LOPO-leaders\\
    \\
    \includegraphics[width=0.3\linewidth]{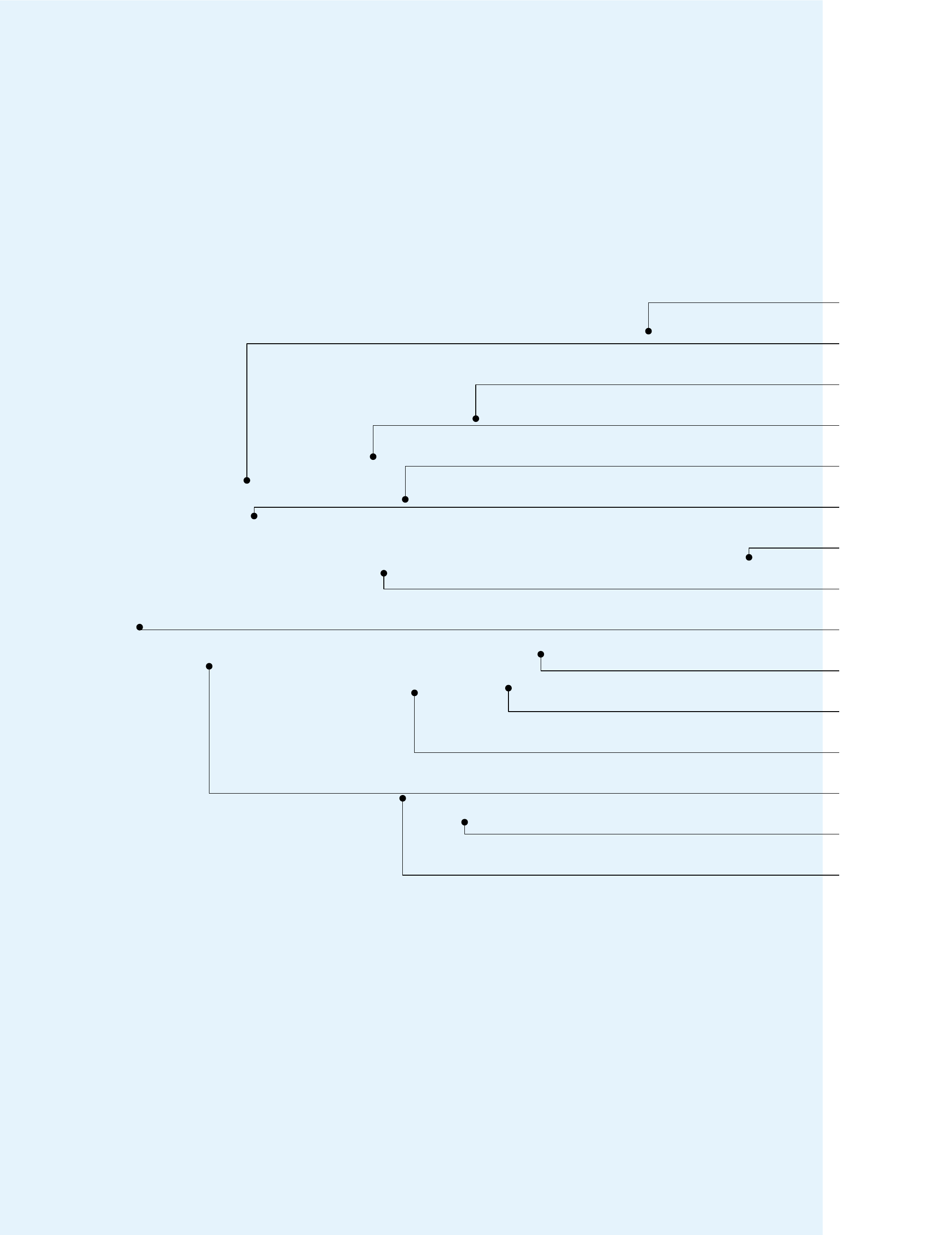} &
    \includegraphics[width=0.3\linewidth]{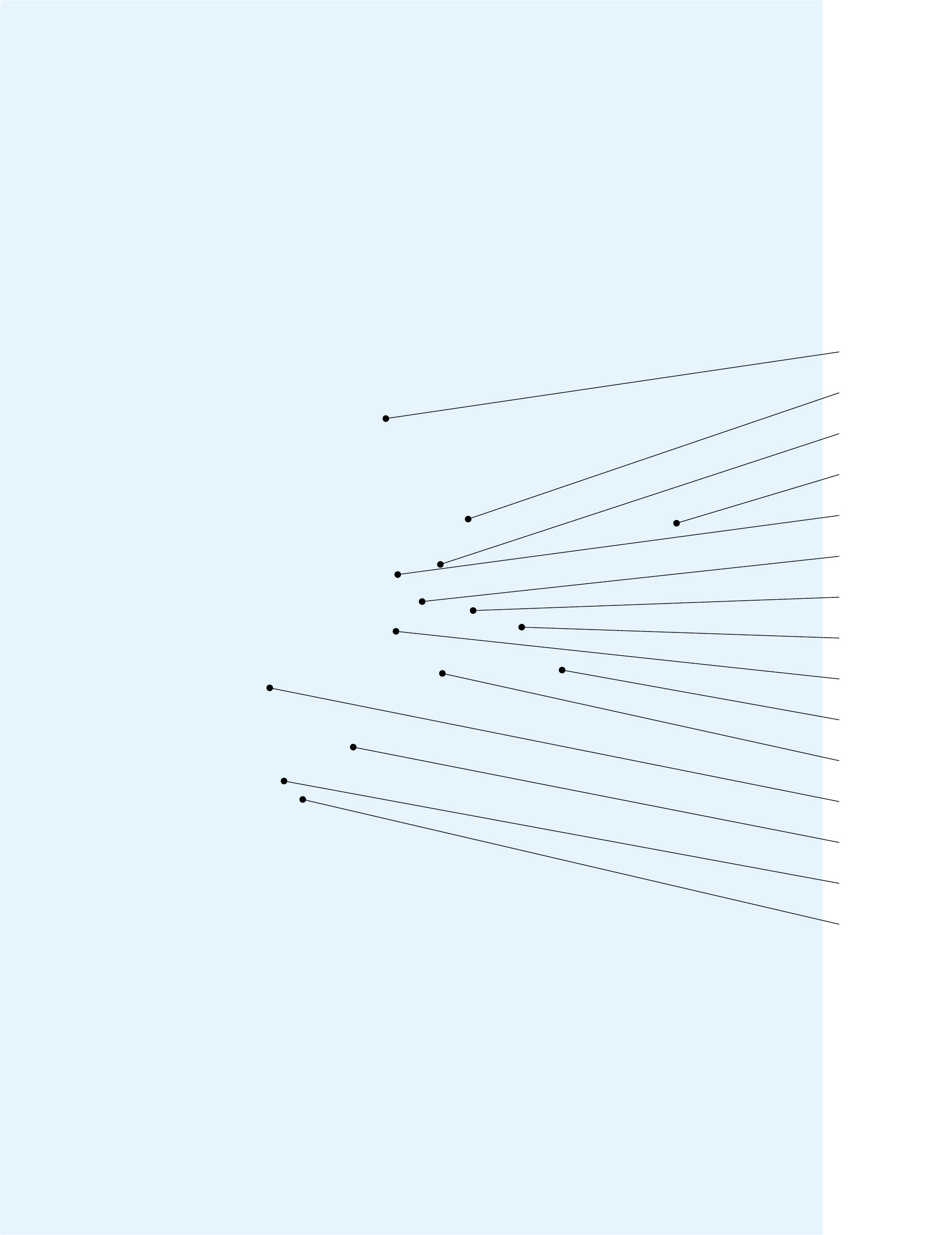}\\
    \LPO-leaders & \LS-leaders\\
  \end{tabular}
\end{center}
\caption{Instances presented as examples next to the
   personal preference questions.}
\label{fig:stimuli:questionnair}
\end{figure}

\begin{figure}[htbp]
  \centering

\newcommand{\scaleA}{0.235}
\centering
\begin{tabular}{lm{3.6cm}cm{3.6cm}cm{3.6cm}}
 & \textit{Dense} &\quad&  \textit{Sparse} &\quad &  \textit{Uniform}   \\
\GPO    & \includegraphics[width=\scaleA\textwidth]{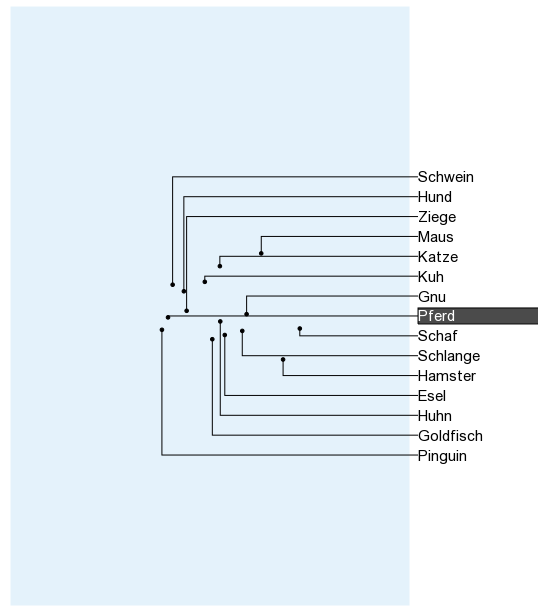} && \includegraphics[width=\scaleA\textwidth]{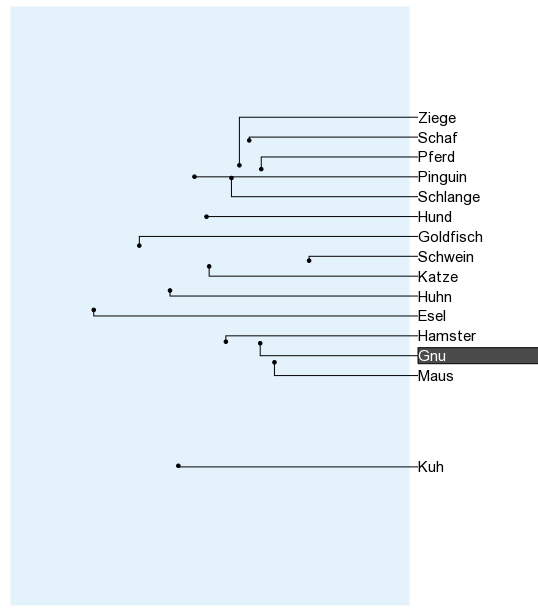} && \includegraphics[width=\scaleA\textwidth]{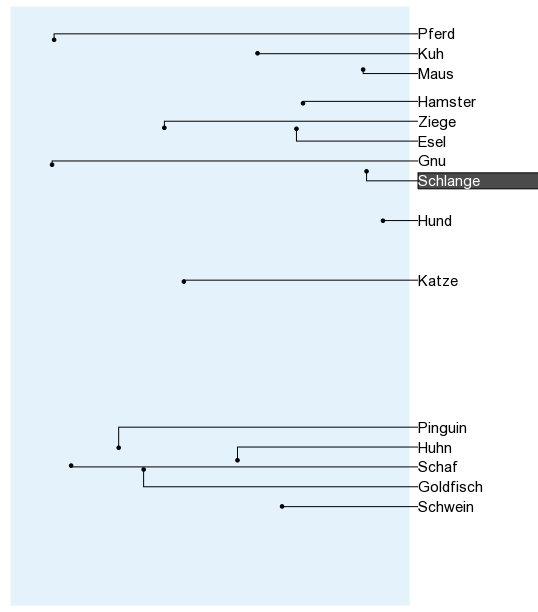} \\
\GS    & \includegraphics[width=\scaleA\textwidth]{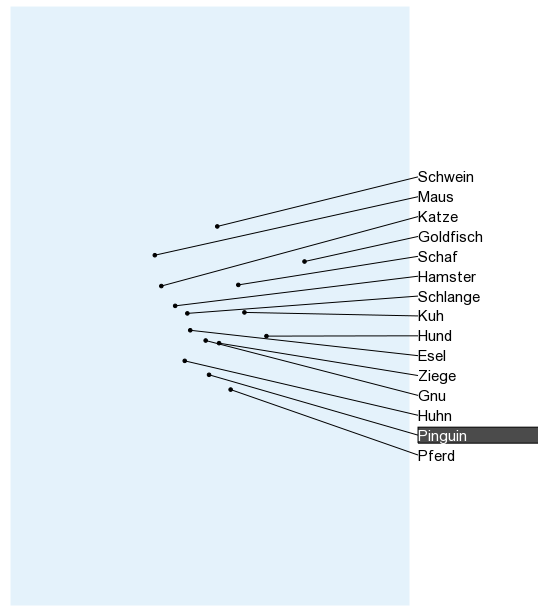} && \includegraphics[width=\scaleA\textwidth]{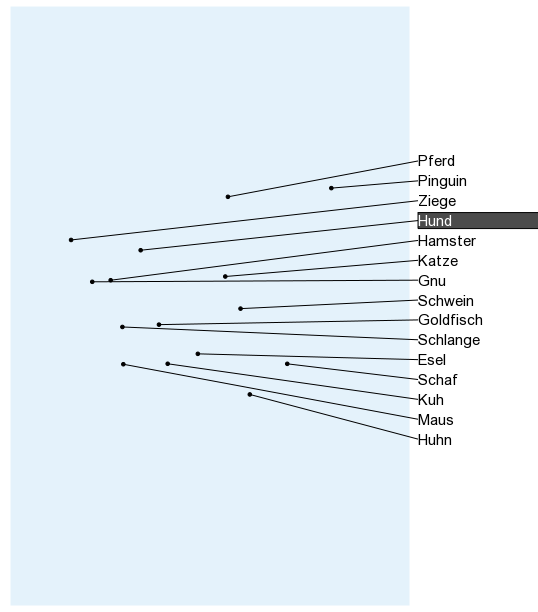} && \includegraphics[width=\scaleA\textwidth]{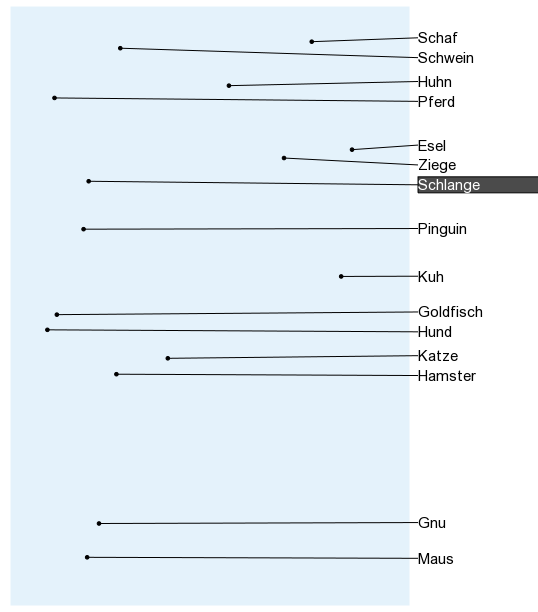} \\
\GDO    & \includegraphics[width=\scaleA\textwidth]{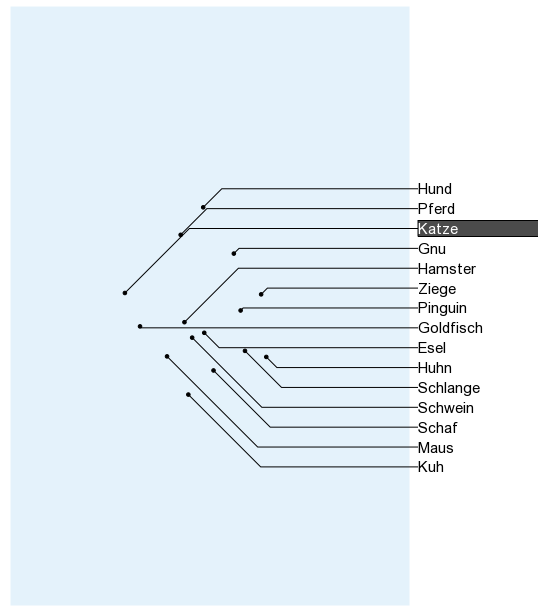} && \includegraphics[width=\scaleA\textwidth]{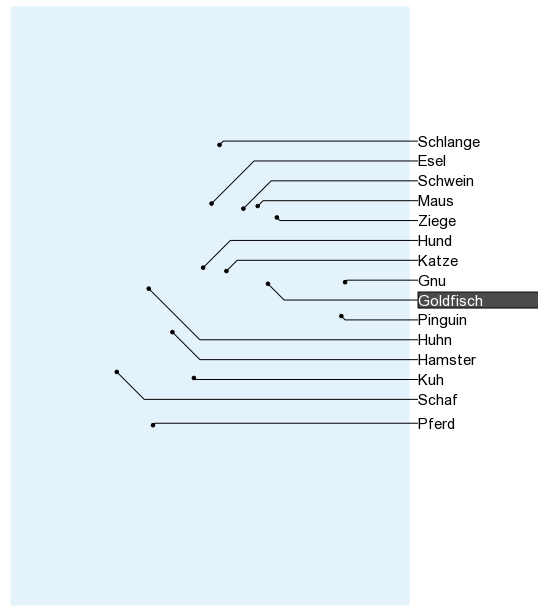} && \includegraphics[width=\scaleA\textwidth]{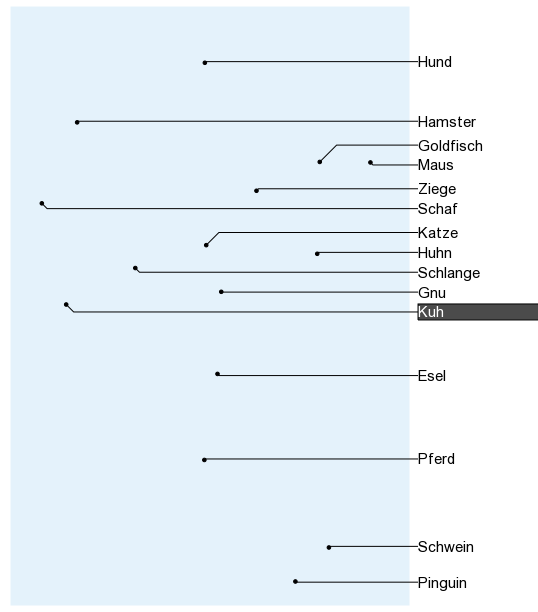} \\
\GOPO    & \includegraphics[width=\scaleA\textwidth]{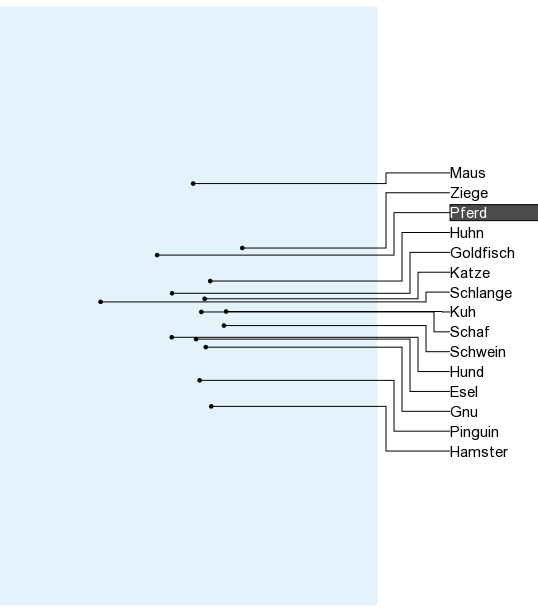} && \includegraphics[width=\scaleA\textwidth]{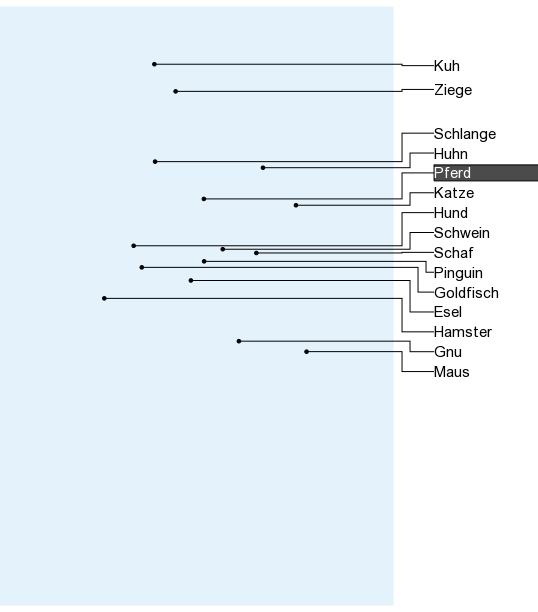} && \includegraphics[width=\scaleA\textwidth]{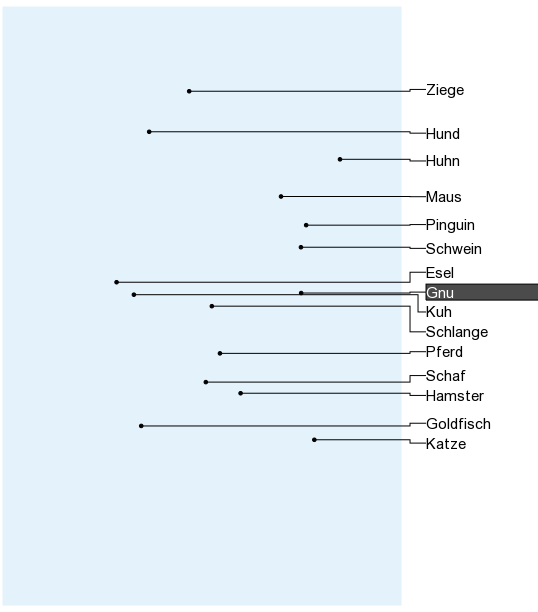} \\
\end{tabular}
  \caption{Example stimuli with 15 sites (small), one for each site distribution and for each leader type. Due to formatting the rectangles enclosing the sites may not have same sizes. In the digital questionnaire they had the same size.}
  \label{table:matrix-small}
\end{figure}

\newcommand{\scaleB}{0.235}

\begin{figure}[htbp]
  \centering

\centering
\begin{tabular}{lm{3.8cm}cm{3.8cm}cm{3.8cm}}
\multicolumn{1}{l}{} & \textit{Dense} &\quad& \textit{Sparse} &\quad & \textit{Uniform}   \\
\GPO    & \includegraphics[width=\scaleB\textwidth]{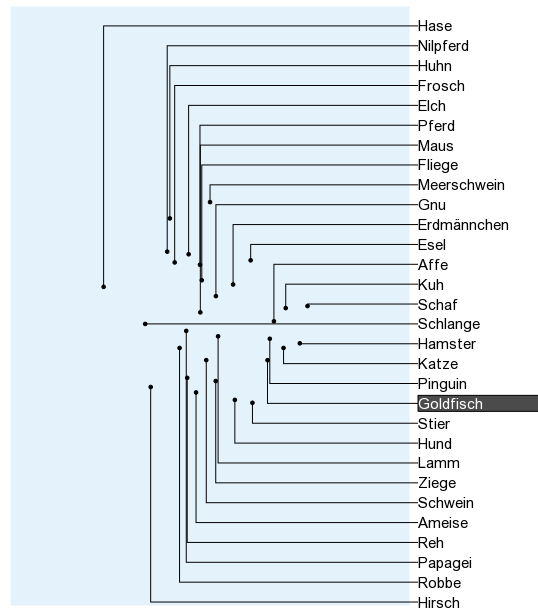} && \includegraphics[width=\scaleB\textwidth]{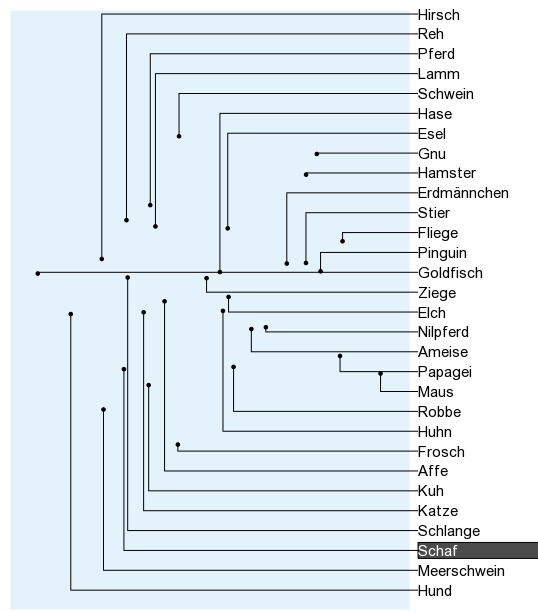} && \includegraphics[width=\scaleB\textwidth]{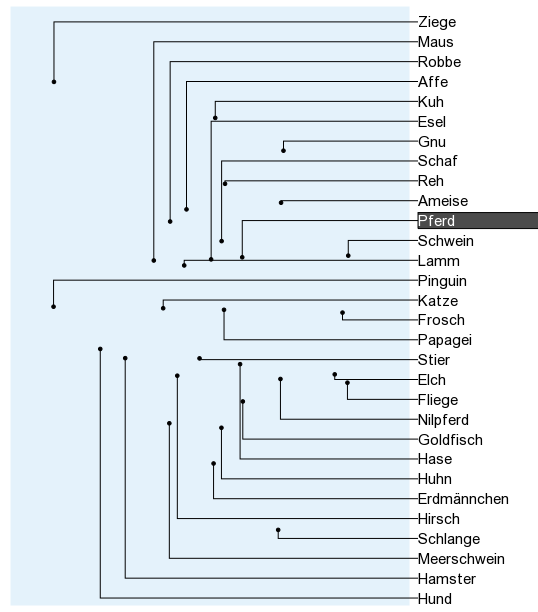} \\
\GS    & \includegraphics[width=\scaleB\textwidth]{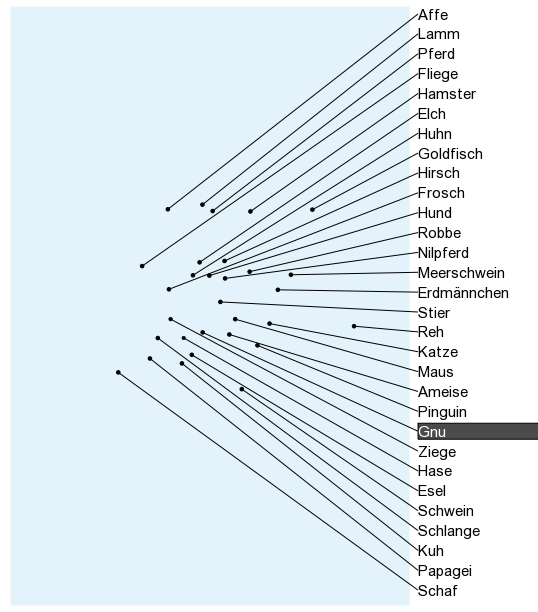} && \includegraphics[width=\scaleB\textwidth]{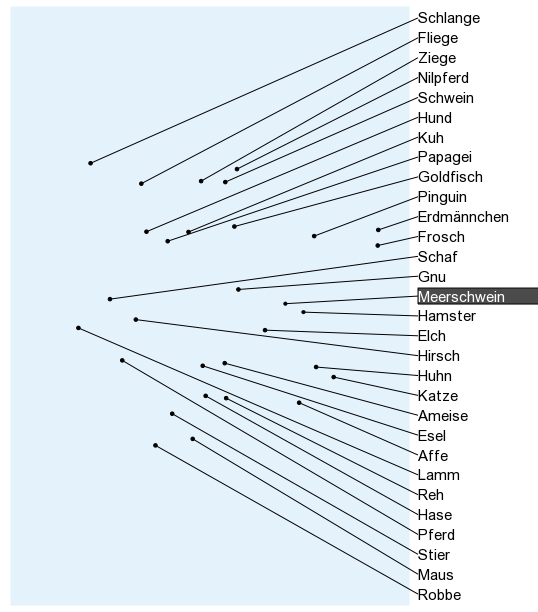} && \includegraphics[width=\scaleB\textwidth]{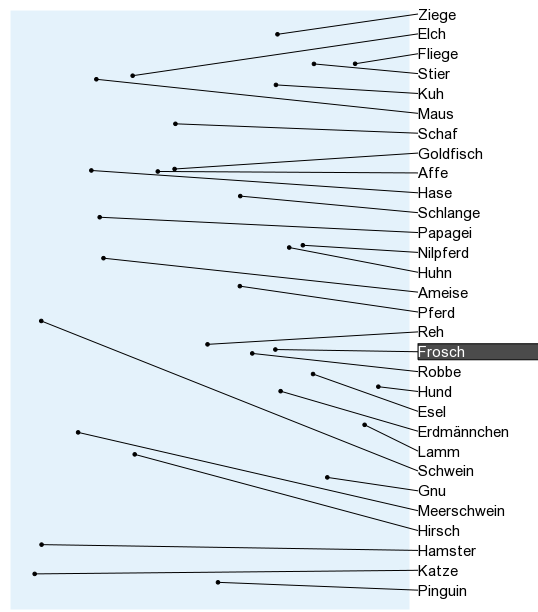} \\
\GDO    & 
\includegraphics[width=\scaleB\textwidth]{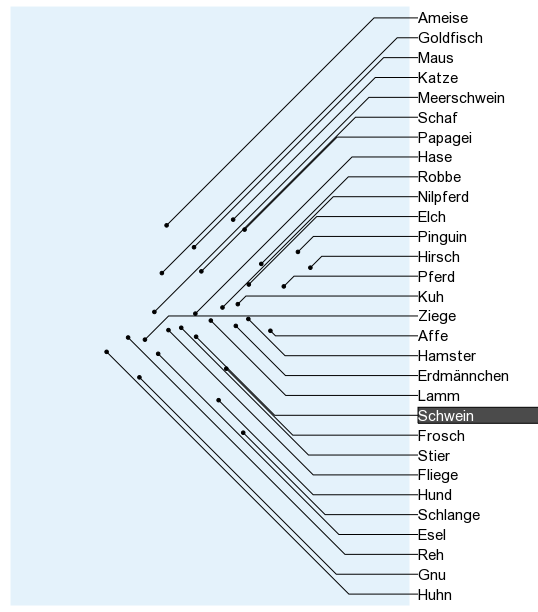} &&
\includegraphics[width=\scaleB\textwidth]{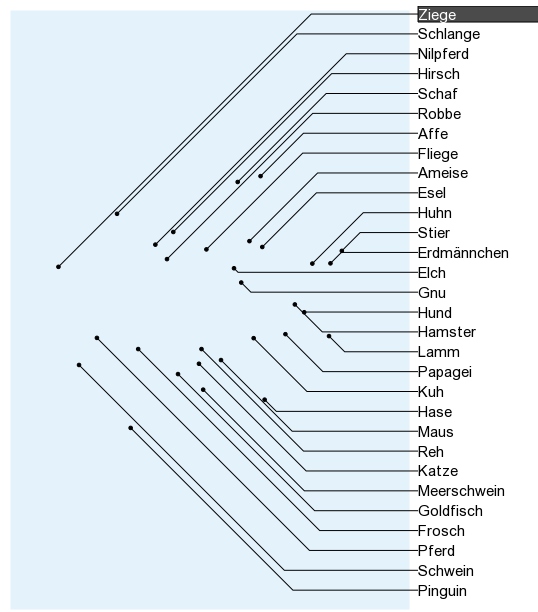} && \includegraphics[width=\scaleB\textwidth]{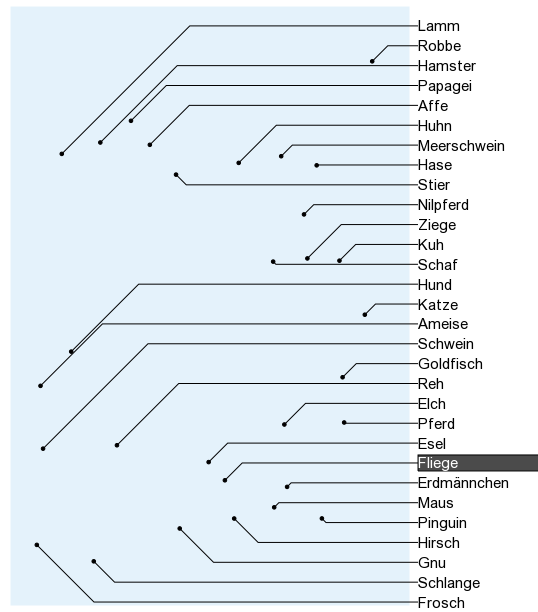} \\
\GOPO    & \includegraphics[width=\scaleB\textwidth]{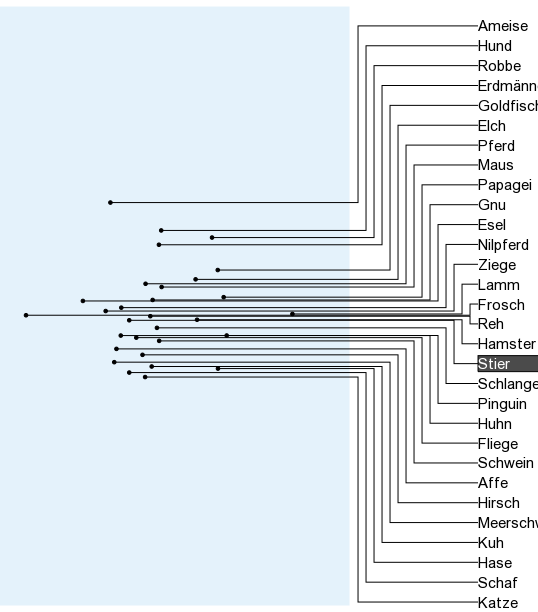} && \includegraphics[width=\scaleB\textwidth]{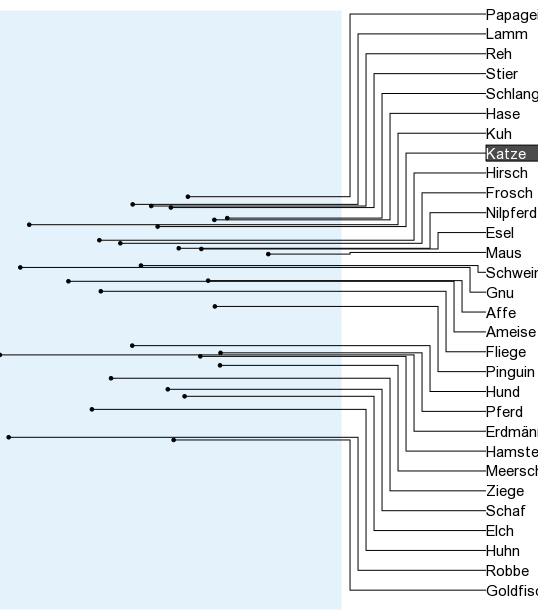} && \includegraphics[width=\scaleB\textwidth]{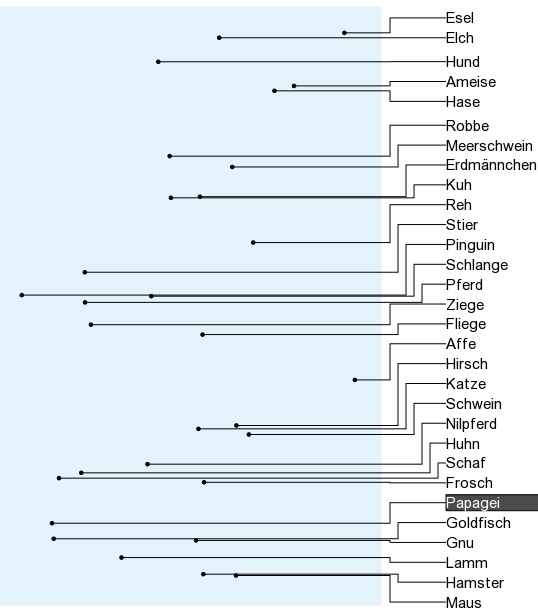} \\
\end{tabular}
  \caption{Example stimuli with 30 sites (large), one for each site distribution and for each leader type. Due to formatting the rectangles enclosing the sites may not have same sizes. In the digital questionnaire they had the same size.}
  \label{table:matrix-large}
\end{figure}

\end{document}